\documentclass[superscriptaddress,aps,prl,twocolumn]{revtex4-2}

\usepackage{amsmath,amssymb,braket}
\usepackage{amsfonts}
\usepackage[dvipdfmx]{graphicx}
\usepackage{color}
\usepackage{bm}
\usepackage{tabularx}
\usepackage[normalem]{ulem}

\usepackage{xcolor}

\newcommand{\sectionprl}[1]{{\em #1}\/.---}

\usepackage[hypertexnames=false]{hyperref}
\hypersetup{
    colorlinks=true,
    linkcolor=blue,
    citecolor=blue,
    urlcolor=blue,
    breaklinks=true
}

\begin{document}

\title{Dissipation anomaly in gradient-driven nonequilibrium steady states}

\date{\today}
\author{Hiroyoshi Nakano}
\affiliation{Institute for Solid State Physics, University of Tokyo, 5-1-5, Kashiwanoha, Kashiwa 277-8581, Japan}

\author{Yuki Minami}
\affiliation{Faculty of Engineering, Gifu University, Yanagido, Gifu 501-1193, Japan}

\date{\today}

\begin{abstract}
Dissipation anomaly—the persistence of finite energy dissipation in the inviscid limit—is a hallmark of turbulence, sometimes regarded as the ``zeroth law'' of turbulent flows.  
Here, we demonstrate that this phenomenon is not exclusive to turbulence.  
Using fluctuating hydrodynamics, we show that a simple gradient-driven nonequilibrium steady state, in which a fluid is subjected to a constant scalar gradient but remains macroscopically quiescent, also exhibits dissipation anomaly.  
Direct numerical simulations and self-consistent mode-coupling theory reveal that the anomaly originates from giant, long-range nonequilibrium fluctuations amplified by the imposed gradient.  
While linear theory predicts a divergent dissipation in the inviscid limit, nonlinear mode coupling regularizes the divergence, yielding a finite anomalous dissipation.  
Our findings identify a new, non-turbulent arena for dissipation anomaly and establish the interplay between thermal noise and nonequilibrium driving as a fundamental route to singular behavior in hydrodynamics.
\end{abstract}

\maketitle

\sectionprl{Introduction}
Dissipation anomaly is a central concept of three-dimensional turbulence~\cite{Frisch1995-rs}, so fundamental that it is sometimes referred to as the ``zeroth law of turbulence.''  
It asserts that, in fully developed turbulence, the mean energy dissipation rate per unit mass $\dot{\varepsilon} := \nu_0 \langle |\nabla \bm{v}|^2 \rangle$
remains finite in the inviscid limit:
\begin{align}
    \lim_{\nu_0 \to 0} \dot{\varepsilon}
    = \lim_{\nu_0 \to 0} \nu_0 \langle |\nabla \bm{v}|^2 \rangle
    = \mathrm{const},
\end{align}
where $\nu_0$ denotes the kinematic viscosity and $\bm{v}$ the velocity field.  
This property entails that the velocity field cannot remain smooth as $\nu_0 \to 0$:  
the mean-squared velocity gradients $\langle |\nabla \bm{v}|^2 \rangle$ must diverge as $1/\nu_0$ to sustain a finite dissipation rate.

%Since the development of these singularities in turbulence was highlighted by Kolmogorov~\cite{Kolmogorov1941-gz, Kolmogorov1941-kx} and Onsager~\cite{Onsager1949-ze}, understanding the behavior of the Navier--Stokes equations in the inviscid limit has been a central challenge in both physics and mathematics.
Since the development of such singularities in turbulence was recognized by Kolmogorov~\cite{Kolmogorov1941-gz, Kolmogorov1941-kx} and Onsager~\cite{Onsager1949-ze}, understanding the behavior of the Navier--Stokes equations in the inviscid limit has remained a central challenge in both physics and mathematics.
Indeed, from a mathematical viewpoint, this phenomenon reflects the rich and complex solution structure of the Euler and Navier--Stokes equations~\cite{De-Lellis2014-pw, Constantin1994-qf, Isett2018-nh}, and is directly linked to the Millennium Prize Problem on Navier--Stokes existence and smoothness.  
From a physical perspective, resolving the spatial structure emerging in this singular limit through direct numerical simulations has been a major challenge for decades~\cite{Sreenivasan1998-sb, Kaneda2003-ni, Yeung2012-fo, Iyer2025-kb}.

Despite this broad interdisciplinary attention, research on inviscid-limit behavior has been almost exclusively confined to turbulent systems.  
Yet, physics offers a rich landscape of nonequilibrium fluid states beyond turbulence---such as fluids under uniform shear or a constant temperature gradient---raising a natural question:  
Must a system be turbulent to develop such singular behavior in the inviscid limit?

In this Letter, we address this question for the first time, demonstrating that both anomalous dissipation and the accompanying singular spatial structures can arise in systems fundamentally different from turbulence.  
Our focus is a general class of systems---fluids driven into nonequilibrium steady states (NESSs) by a constant external gradient.  
This class encompasses, for example, a single-component fluid under a uniform temperature gradient or a binary mixture subject to a concentration gradient.  
Unlike turbulent flows, these systems are macroscopically quiescent, with a vanishing mean velocity, $\langle \bm{v}\rangle = \bm{0}$.  
Consequently, within conventional deterministic hydrodynamics~\cite{Landau1987-eu}, they are not expected to exhibit the dissipation anomaly or the associated singular behavior.

Our central idea is to generalize the notion of the inviscid limit within the framework of fluctuating hydrodynamics~\cite{De_Zarate2006-xw, Das2011-ao}.  
This framework extends conventional hydrodynamics by incorporating thermal fluctuations arising from microscopic molecular motions.  
The inclusion of such fluctuations is particularly essential for the gradient-driven NESSs considered here:  
an external gradient is known to anomalously amplify these thermal motions, giving rise to giant, long-range nonequilibrium fluctuations~\cite{De_Zarate2006-xw, Dorfman1994-cl, Bedeaux2015-lu, Sengers2024-gz, Ronis1982-uk, Lutsko1985-zb, Brogioli2001-ct, Donev2011-hf, Kirkpatrick2015-ai, Peraud2016-fd, Peraud2017-xt, Kirkpatrick2021-rv, Nakano2022-kv, Nakano2025-zw, Bussoletti2025-qh, Bussoletti2025-jd, Vailati1997-ez, Takacs2011-gg, Vailati2011-rz, Vailati2012-yz}.

Building on this established understanding, we examine the inviscid limit of the fluctuating hydrodynamic equations under a fixed external gradient.  
Our analysis yields two main results.  
First, anomalous dissipation arises even in this simple gradient-driven NESS.
Second, its fundamental origin is traced to the long-range nonequilibrium fluctuations.

These conclusions are established through two complementary approaches: direct numerical simulations of fluctuating hydrodynamics and a self-consistent mode-coupling theory.
Our numerical simulations, in particular, employ recently developed techniques~\cite{BalboaUsabiaga2012-sh, Delong2013-fh, Srivastava2023-nx, Garcia2024-nq} that enable us to effectively access the inviscid limit and provide crucial evidence for the anomaly.

Our findings establish a new, non-turbulent setting for the dissipation anomaly and connect with recent efforts to elucidate the role of thermal fluctuations in the singular limits of hydrodynamics.  
Pioneer studies have primarily advanced along two directions: the influence of thermal fluctuations on the dissipation range of turbulence~\cite{McMullen2022-xb, Bell2022-en, Eyink2022-fc, Bandak2022-rk, Bandak2024-al, Ishan2025-fw} and the inviscid-limit dynamics of the Karder-Parisi-Zhang (KPZ) equation~\cite{Rodriguez-Fernandez2022-qw, Cartes2022-ur, Fontaine2023-nr} and thermal equilibrium systems~\cite{Gosteva2025-cq}.  
The present work adds a new dimension by demonstrating that the interplay between thermal fluctuations and a nonequilibrium drive alone is sufficient to produce dissipation anomaly, even in the absence of a turbulent cascade.

\sectionprl{Model}
We consider a scalar field $\psi(\bm{r},t)$ advected by an incompressible velocity field $\bm{v}(\bm{r},t)$.
The dynamics is governed by the fluctuating Navier--Stokes and convection--diffusion equations,  
\begin{align}
\frac{\partial \bm{v}}{\partial t} = -\frac{1}{\rho_0} \nabla p + \nu_0 \nabla^2 \bm{v} - \nabla \cdot \bm{\Pi}_{\mathrm{ran}}, \label{eq:sto_ns} \\
\frac{\partial \psi}{\partial t} + \nabla \cdot (\psi \bm{v}) = D_0 \nabla^2 \psi - \nabla \cdot \bm{J}_{\mathrm{ran}}.\label{eq:sto_cd}
\end{align}
These equations are subjected to the incompressibility condition $\nabla \cdot \bm{v} = 0$.
Here, $\rho_0$ is the constant mass density, $\nu_0$ is the kinematic viscosity, and $D_0$ is the diffusion coefficient for $\psi$.
The pressure $p$ is determined so as to enforce $\nabla \cdot \bm{v} = 0$.
We omit the nonlinear advective term $(\bm{v}\!\cdot\!\nabla)\bm{v}$ from the Navier--Stokes equation [Eq.~(\ref{eq:sto_ns})] to suppress the development of turbulence in the inviscid limit.  
The stochastic fluxes $\bm{\Pi}_{\mathrm{ran}}$ and $\bm{J}_{\mathrm{ran}}$ represent thermal noise,  
whose correlations obey the fluctuation–dissipation theorem:
\begin{align}
\begin{aligned}
    \left \langle \Pi_{\mathrm{ran}}^{ab}(\bm{r},t)\,
        \Pi_{\mathrm{ran}}^{cd}(\bm{r}',t') \right \rangle
    &= 2\frac{k_B T}{\rho_0}\nu_0
       \delta(\bm{r}-\bm{r}')\delta(t-t') \\
    &\quad \times \!
    \Big(\delta^{ac}\delta^{bd}+\delta^{ad}\delta^{bc}
          -\tfrac{2}{d}\delta^{ab}\delta^{cd}\Big),\\[2pt]
    \left\langle J_{\mathrm{ran}}^{a}(\bm{r},t)\,
        J_{\mathrm{ran}}^{b}(\bm{r}',t') \right\rangle
    &= 2 k_B T \chi D_0\, \delta^{ab}\delta(\bm{r}-\bm{r}')\delta(t-t'),
\end{aligned}
\end{align}
where $k_B$ is the Boltzmann constant and $T$ is the temperature.
The parameter $\chi$ denotes the static susceptibility, defined through the equilibrium distribution of the scalar field,  
$P[\psi] \propto \exp[-F[\psi]/(k_B T)]$,  
with the quadratic free energy functional  
$F[\psi] = (1/2\chi) \int d^d\bm{r}\,\psi(\bm{r})^2$.

This model provides, for instance, a minimal description of a binary fluid mixture, where $\psi$ represents the concentration of one component~\cite{Donev2011-hf, Donev2014-jy, Donev2015-tm}.
Note that, since Eq.~(\ref{eq:sto_ns}) for the velocity field is linear, $\bm{v}$ can be obtained exactly as a response proportional to the random stress $\bm{\Pi}_{\mathrm{ran}}$.
The system nevertheless remains nonlinear through the coupling term $\nabla\!\cdot\!(\psi \bm{v})$ in Eq.~(\ref{eq:sto_cd}), which constitutes the central focus of this Letter.

We drive the system into a NESS by imposing a constant mean gradient $G$ on the scalar field, $\langle \psi(\bm{r}) \rangle = \psi_0 + G y$.  
The dynamics of fluctuations about this mean profile, $\delta\psi := \psi - \langle \psi \rangle$, are then governed by  
\begin{align}
\frac{\partial (\delta\psi)}{\partial t} + G v_y + \bm{v}\!\cdot\!\nabla (\delta\psi)
&= D_0 \nabla^2 (\delta\psi) - \nabla\!\cdot\!\bm{J}_{\mathrm{ran}}, \label{eq:sto_cd_fluct}
\end{align}
where $v_y$ is the $y$ component of the velocity field.

It should be noted that, in fluctuating hydrodynamics, the spatial grid spacing $a_{\mathrm{uv}}$ used in numerical simulations serves as a parameter of the theory, on the same footing as $\nu_0$ and $D_0$.
The grid spacing is related to the ultraviolet wavenumber cutoff $\Lambda = 2\pi / a_{\mathrm{uv}}$ in analytic calculations, and its dependence and physical meaning are discussed in Ref.~\cite{Nakano2025-tj}.  
Throughout this work, we fix $a_{\mathrm{uv}}$ and focus on the dependence of physical quantities on the transport coefficients.

%---------------flguire------------------
\begin{figure}[t]
\begin{center}
\includegraphics[scale=1.0]{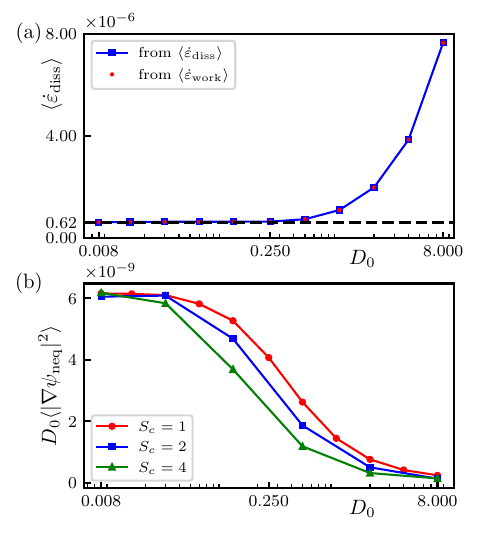}
\end{center}
\vspace{-0.5cm}
\caption{
Numerical results for the anomalous dissipation in our model. The parameters are set to $\rho_0=k_B T=1$, $\chi=0.01$, $L=2048$, $a_{\rm uv}=32$, and $G=0.2/L$.
(a) The noise-averaged dissipation rate $\langle \dot{\varepsilon}_{\rm diss} \rangle$ as a function of $D_0$ for a fixed Schmidt number $S_c = 1$.
The work rate $\langle \dot{\varepsilon}_{\rm work} \rangle$, calculated independently, is overlaid as the red symbols.
(b) The quantity $D_0 \langle |\nabla \delta \psi_{\rm neq}|^2 \rangle$ as a function of $D_0$ for several Schmidt numbers.
}
\label{fig1}
\end{figure}
%---------------flguire------------------

\sectionprl{Dissipation anomaly in the inviscid limit}
The central quantity in our analysis is the energy dissipation rate per unit volume, $\dot{\varepsilon}_{\mathrm{diss}}$.  
This quantity is obtained by analyzing the balance of the total free energy $F_{\rm tot} = \int d^d\bm{r} \left(\psi^2/2\chi + \rho_0 \bm{v}^2/2 \right)$
within the framework of irreversible thermodynamics~\cite{degroot2013-ib, chaikin1995-pr}.  
This analysis shows that the dissipation rate can be generally expressed as the product of an irreversible flux and its conjugate thermodynamic force.  
For our model, this relation takes the explicit form~\footnote{See Supplemental Material (SM).}  
\begin{align}
\dot{\varepsilon}_{\mathrm{diss}}
    = -\frac{1}{V_d} \int_{V_d} d^d\bm{r}\,
      \bm{J}_{\mathrm{irr}} \cdot \nabla\!\left(\frac{\psi}{\chi}\right),
    \label{eq:dissipation_rate_general}
\end{align}
where $\bm{J}_{\mathrm{irr}} := -D_0 \nabla \psi + \bm{J}_{\mathrm{ran}}$ is the irreversible flux of $\psi$.  
Note that the velocity field $\bm{v}$ remains in equilibrium and therefore does not contribute to entropy production.

In this Letter, we focus on the noise-averaged dissipation rate $\langle \dot{\varepsilon}_{\mathrm{diss}} \rangle$ in the nonequilibrium steady state.  
To identify the contribution relevant to the dissipation, we decompose the scalar fluctuations as $\delta\psi = \delta\psi_{\mathrm{eq}} + \delta\psi_{\mathrm{neq}}$,  
where $\delta\psi_{\mathrm{eq}}$ denotes the equilibrium fluctuations present even at $G=0$, and $\delta\psi_{\mathrm{neq}}$ represents the additional nonequilibrium component induced by the gradient $G$~\footnote{
See End Matter.
}.  
As shown in the Supplemental Material (SM)~\footnotemark[1], $\delta\psi_{\mathrm{eq}}$ does not contribute to $\langle \dot{\varepsilon}_{\mathrm{diss}} \rangle$.  
Substituting the above decomposition into Eq.~(\ref{eq:dissipation_rate_general}) and taking the noise average yields  
\begin{align}
\langle \dot{\varepsilon}_{\mathrm{diss}} \rangle
    = \frac{D_0}{\chi} G^2
      + \frac{D_0}{\chi} \left\langle |\nabla \delta \psi_{\mathrm{neq}}|^2 \right\rangle.
    \label{eq:vareps_dissp}
\end{align}

We now present the central result of this Letter: the dissipation rate $\langle \dot{\varepsilon}_{\mathrm{diss}} \rangle$ exhibits dissipation anomaly, as shown in Fig.~\ref{fig1}.  
We examine the inviscid limit by taking the bare transport coefficients $\nu_0$ and $D_0$ to zero while keeping the mean gradient $G$ fixed, together with all other parameters $\rho_0$, $k_B T$, $\chi$, $a_{\mathrm{uv}}$, and $L$~\footnote{
Based on the dimensional analysis, we find that this limit $\nu_0, D_0 \!\to\! 0$ corresponds to the regime where the dimensionless measure of thermal fluctuations, $k_B T / (\rho_0 a_{\mathrm{uv}}^{d-2} D_0^2)$, becomes large.
This can be verified by comparing the first and second terms of Eq.~(\ref{eq:vareps_dissp_lin}).
}.
We show below that in this limit, $\langle \dot{\varepsilon}_{\mathrm{diss}} \rangle$ approaches a finite value,
\begin{align}
\lim_{\nu_0, D_0 \to 0} \langle \dot{\varepsilon}_{\mathrm{diss}} \rangle
    = \lim_{\nu_0, D_0 \to 0} \frac{D_0}{\chi} 
      \left\langle |\nabla \delta \psi_{\mathrm{neq}}|^2 \right\rangle
    = \dot{\varepsilon}_{\ast} > 0.
\end{align}

To demonstrate this, we perform direct numerical simulations of the full nonlinear equations [Eqs.~(\ref{eq:sto_ns}) and (\ref{eq:sto_cd_fluct})]~\footnotemark[2].  
The simulations employ the recently developed method~\cite{Srivastava2023-nx, Garcia2024-nq}, combining real-space discretization on a staggered grid~\cite{BalboaUsabiaga2012-sh} with low-storage third-order Runge--Kutta (RK3) time integration~\cite{Delong2013-fh}. 
Figure~\ref{fig1}(a) shows $\langle \dot{\varepsilon}_{\mathrm{diss}} \rangle$ as a function of $D_0$ at fixed Schmidt number $S_c = \nu_0 / D_0 = 1$.  
As $D_0$ decreases, the data clearly exhibit a plateau at a finite value $\dot{\varepsilon}_{\ast}$, providing direct evidence for the dissipation anomaly.
We comment that for computational efficiency, we simulated a two-dimensional system, although the theoretical analysis below indicates that the conclusions remain qualitatively valid in three dimensions.

To sustain the finite dissipation, the mean-squared gradient $\langle |\nabla \delta \psi_{\mathrm{neq}}|^2 \rangle$ must diverge.  
We test this by directly evaluating $D_0 \langle |\nabla \delta \psi_{\mathrm{neq}}|^2 \rangle$.  
This quantity is obtained from the dissipation rate via  
$D_0 \langle |\nabla \delta \psi_{\mathrm{neq}}|^2 \rangle = \chi \langle \dot{\varepsilon}_{\mathrm{diss}} \rangle - D_0 G^2$,  
a rearrangement of Eq.~(\ref{eq:vareps_dissp}).  
As shown in Fig.~\ref{fig1}(b) for several Schmidt numbers $S_c$,  
$D_0 \langle |\nabla \delta \psi_{\mathrm{neq}}|^2 \rangle$ converges to a finite constant in the inviscid limit.  
This implies that $\langle |\nabla \delta \psi_{\mathrm{neq}}|^2 \rangle$ diverges, scaling as $1/D_0$.
Crucially, this divergence signifies the emergence of singular spatial structures in the inviscid limit.
This conclusion is further reinforced by our numerical finding that the fluctuation amplitude $\sqrt{\langle |\delta \psi_{\mathrm{neq}}|^2 \rangle}$ also diverges, as detailed in the End Matter (EM)~\footnotemark[2].
% Remarkably, the limiting value is independent of $S_c$, a feature explained by the theory below.

We further show that the anomalous dissipation is equivalent to the persistence of a finite mean flux of $\psi$.  
In the steady state, the dissipated energy must be balanced by the work done on the system to maintain the constant gradient, $\langle \dot{\varepsilon}_{\mathrm{work}} \rangle$, which is expressed as $\langle \dot{\varepsilon}_{\mathrm{work}} \rangle = -(G/\chi)\langle J_\psi^y\rangle$, where $\bm{J}_\psi := \psi\bm{v} - D_0\nabla\psi + \bm{J}_{\mathrm{ran}}$ is the total flux of $\psi$~\footnotemark[1].  
The steady-state energy balance $\langle \dot{\varepsilon}_{\mathrm{work}} \rangle = \langle \dot{\varepsilon}_{\mathrm{diss}} \rangle$
then gives a direct relation $\langle J_\psi^y\rangle = -(\chi/G)\langle \dot{\varepsilon}_{\mathrm{diss}}\rangle$, showing that a finite dissipation necessarily implies a finite mean flux,
\begin{align}
    \lim_{\nu_0,D_0\to0}\langle J_\psi^y\rangle = -\frac{\chi}{G}\dot{\varepsilon}_\ast < 0.
\end{align}
In the simulations, we independently compute $\langle \dot{\varepsilon}_{\mathrm{work}}\rangle$ from the flux of $\psi$ and confirm that it matches the dissipation rate almost perfectly, as shown by the overlaid symbols in Fig.~\ref{fig1}(a).  
This excellent agreement serves as a consistency check of our results.

%---------------flguire------------------
\begin{figure}[t]
\begin{center}
\includegraphics[scale=1.0]{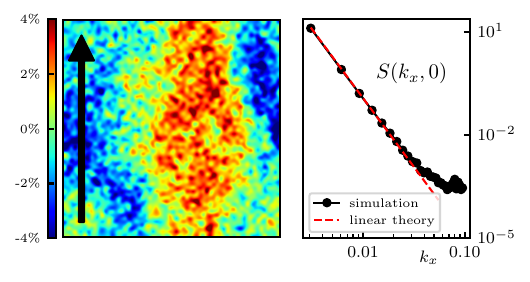}
\end{center}
\vspace{-0.5cm}
\caption{
Numerical confirmation of long-range correlations in the linear regime ($D_0=2.0$).
Simulation parameters are otherwise the same as in Fig.~\ref{fig1}.
(a) A typical real-space snapshot of the fluctuation field $\delta\psi$.
The black arrow is the direction of the imposed gradient $G$.
(b) The corresponding spatial correlation of the non-equilibrium fluctuations, $S_{\rm neq}(\bm{k})$, plotted along the $k_x$ axis ($k_y=0$).
The numerical data are compared with the prediction from the linearized theory [Eq.~(\ref{eq:stfac_lin})], shown as a dashed red line.
}
\label{fig2}
\end{figure}
%---------------flguire------------------
\sectionprl{Origin of the Anomaly}
The anomalous dissipation originates from giant, long-range nonequilibrium fluctuations, a hallmark of the NESS in our model.  
To elucidate this mechanism, we first analyze the model within the linear approximation.

Within this linear approximation, where the nonlinear advection term in Eq.~(\ref{eq:sto_cd_fluct}) is neglected,  
the spatial correlations of the nonequilibrium fluctuations,  
$S_{\mathrm{neq}}(\bm{k}) := \langle |\tilde{\delta \psi}_{\mathrm{neq}}(\bm{k})|^2\rangle / V_d$,  
are obtained analytically as~\cite{De_Zarate2006-xw}
\begin{align}
    S_{\mathrm{neq}}(\bm{k})
        \approx \frac{k_B T}{\rho_0}
        \frac{G^2}{D_0(\nu_0 + D_0)}
        \frac{|\bm{k}|^2 - k_y^2}{|\bm{k}|^6}.
    \label{eq:stfac_lin}
\end{align}
Here, $\tilde{\delta\psi}_{\mathrm{neq}}(\bm{k})$ is the Fourier transform of $\delta\psi_{\mathrm{neq}}(\bm{r})$, and $V_d$ is the system volume.
This result predicts a characteristic $|\bm{k}|^{-4}$ scaling of $S_{\mathrm{neq}}(\bm{k})$ at small wave numbers, a hallmark of long-range correlations.

These long-range fluctuations are clearly observed in our simulations of the full nonlinear model, which is performed in the large-$D_0$ regime to neglect the nonlinear effects.
Figure~\ref{fig2}(a) shows a representative real-space snapshot of $\delta\psi$, revealing large-scale structures strongly elongated along the direction of the imposed gradient (see SM for movies).
Quantitatively, as shown in Fig.~\ref{fig2}(b), the computed $S_{\mathrm{neq}}(\bm{k})$ agrees closely with the linear theory [Eq.~(\ref{eq:stfac_lin})], including the characteristic $|\bm{k}|^{-4}$ scaling.

The dissipation rate $\langle \dot{\varepsilon}_{\mathrm{diss}} \rangle$ is directly related to $S_{\mathrm{neq}}(\bm{k})$ through
\begin{align}
    \langle \dot{\varepsilon}_{\mathrm{diss}} \rangle
        = \frac{D_0}{\chi} G^2
        + \frac{D_0}{\chi} \int \frac{d^d\bm{k}}{(2\pi)^d} |\bm{k}|^2 S_{\mathrm{neq}}(\bm{k}).
    \label{eq:dissp_and_stfac}
\end{align}

By substituting the linear-theory prediction for $S_{\mathrm{neq}}(\bm{k})$ [Eq.~(\ref{eq:stfac_lin})] into Eq.~(\ref{eq:dissp_and_stfac}), we obtain the dissipation rate within the linear approximation,
\begin{align}
    \langle \dot{\varepsilon}_{\mathrm{diss}} \rangle
        \approx \frac{D_0}{\chi} G^2
        + \frac{k_B T}{\rho_0} \frac{G^2}{\chi (\nu_0 + D_0)}
          \int \frac{d^d\bm{k}}{(2\pi)^d}
          \frac{|\bm{k}|^2 - k_y^2}{|\bm{k}|^4}.
    \label{eq:vareps_dissp_lin}
\end{align}
The second term represents the contribution of the long-range correlations.
Crucially, it is proportional to $1/(\nu_0 + D_0)$, a dependence that prevents $\langle \dot{\varepsilon}_{\mathrm{diss}} \rangle$ from vanishing in the inviscid limit, thereby leading to the anomalous dissipation.

%---------------flguire------------------
\begin{figure}[t]
\begin{center}
\includegraphics[scale=1.0]{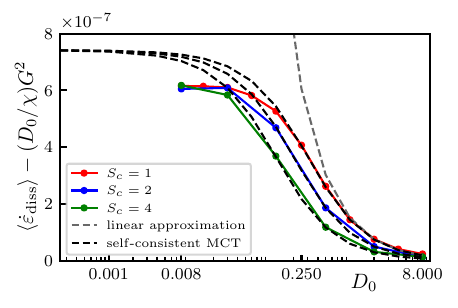}
\end{center}
\vspace{-0.5cm}
\caption{
Quantitative comparison between the numerical simulation results and the theoretical predictions.
Simulation parameters are otherwise the same as in Fig.~\ref{fig1}.
The fluctuation-induced part of the dissipation rate, $\langle\dot{\varepsilon}_{\mathrm{diss}}\rangle - (D_0/\chi)G^2$, is plotted as a function of $D_0$ for several $S_c$.
The numerical data are compared with the predictions of the self-consistent MCT (black dashed curves) and the linearized theory (gray dashed curve).
}
\label{fig3}
\end{figure}
%---------------flguire------------------
\sectionprl{Self-consistent one-loop mode-coupling theory}
The linear approximation indicates that the anomaly originates from long-range correlations but cannot fully capture the inviscid limit, yielding a divergence instead of the finite dissipation observed in our simulations.  
This discrepancy underscores the essential role of the nonlinear advection term.  
We next show that this term regularizes the divergence, yielding a finite anomalous dissipation.

To this end, we employ the framework of mode-coupling theory (MCT)~\cite{Kawasaki1970-vo, Kawasaki1971-am, Bedeaux1974-da, Mazur1974-bs, Pomeau1975-ll}. In this framework~\footnotemark[1], the nonlinear advection renormalizes the bare diffusion coefficient $D_0$ into a scale-dependent, renormalized coefficient $D_R(\bm{k})$.
The system’s dynamics can thus be effectively described by a linear equation with $D_R(\bm{k})$ in place of $D_0$.
The problem thus reduces to finding an analytic expression for $D_R(\bm{k})$. 
Once an expression for $D_R(\bm{k})$ is obtained, the dissipation rate $\langle \dot{\varepsilon}_{\rm dissp}\rangle$ is given as a natural extension of the linear theory result [Eq.~(\ref{eq:vareps_dissp_lin})], which now depends on $D_R(\bm{k})$:
\begin{align}
    \langle \dot{\varepsilon}_{\rm diss}\rangle \approx \frac{D_0}{\chi} G^2 + \frac{k_B T}{\rho_0} \frac{G^2}{\chi} \int \frac{d^d\bm{k}}{(2\pi)^d} \frac{1}{\nu_0 + D_R(\bm{k})} \frac{|\bm{k}|^2-k_y^2}{|\bm{k}|^4}.
    \label{eq:wr_scmct_final}
\end{align}

In this Letter, we determine $D_R(\bm{k})$ by a self-consistent treatment of the one-loop diagram~\cite{Bedeaux1974-da, Mazur1974-bs, Pomeau1975-ll}, which is consistent with the renormalization group approach by Forster \textit{et al.}~\cite{Forster1977-lr}.
The detailed calculation is presented in the SM~\footnotemark[1], and the resulting analytical expression is
\begin{align}
    D_R(\bm{k}) = \frac{D_0 - \nu_0}{2}
        + \sqrt{\left(\frac{D_0 + \nu_0}{2}\right)^2 + \Delta(\bm{k}) } ,
    \label{eq:DR_final}
\end{align}
where $\Delta(\bm{k}) = \bigl(k_B T(d-1)S_{d-1}/\rho_0d(2\pi)^d\bigr) 
        \int_{k}^{\Lambda} q^{d-3}\,dq$.
%The crucial feature of this result is that in the inviscid limit, $D_R(\bm{k})$ converges to a finite, non-zero value $\sqrt{\Delta(\bm{k})}$. Consequently, $\langle \dot{\varepsilon}_{\rm dissp}\rangle$ calculated from Eq.~(\ref{eq:wr_scmct_final}) also converges to a finite value, thus resolving the divergence predicted by the linear approximation.
The crucial feature of this result is that, in the inviscid limit, $D_R(\bm{k})$ approaches a finite, non-zero value $\sqrt{\Delta(\bm{k})}$.  
Consequently, $\langle \dot{\varepsilon}_{\mathrm{diss}}\rangle$ calculated from Eq.~(\ref{eq:wr_scmct_final}) also converges to a finite value, resolving the divergence predicted by the linear approximation.

We quantitatively compare the MCT prediction with our simulation results.  
Figure~\ref{fig3} shows the fluctuation-induced part of the dissipation rate,  
$\langle\dot{\varepsilon}_{\mathrm{diss}}\rangle - (D_0/\chi)G^2$, as a function of $D_0$.  
As shown in the figure, the MCT [Eq.~(\ref{eq:wr_scmct_final}) combined with Eq.~(\ref{eq:DR_final})] provides an excellent description of the simulation data over a wide range of $D_0 \ge 0.12$.  
In contrast, the linear theory is accurate only for $D_0 \gtrsim 2.0$.

Furthermore, the MCT correctly captures the convergence to a finite value in the inviscid limit, even though its quantitative accuracy is not perfect.  
The predicted anomalous dissipation agrees reasonably well with the simulation, with a discrepancy of about $20\%$.  
Such quantitative agreement represents a nontrivial success of the self-consistent one-loop MCT, confirming that the theory captures the essential physics of the anomalous dissipation.
We note, however, that achieving fully quantitative predictions in the inviscid limit remains an open challenge.  
This would likely require incorporating higher-order corrections beyond the one-loop level or adopting nonperturbative approaches such as the functional renormalization group~\cite{Gosteva2025-cq}.

\sectionprl{Summary and discussion}
In this Letter, we have demonstrated the existence of dissipation anomaly in a simple nonequilibrium steady state (NESS)—a fluid driven by a uniform scalar gradient.
We have further shown that this anomaly is directly linked to the divergence of the nonequilibrium fluctuation gradients $\nabla \delta \psi_{\rm neq}$, suggesting the emergence of singular spatial structures in the inviscid limit.
This finding was established numerically, leveraging recent computational advances~\cite{BalboaUsabiaga2012-sh, Delong2013-fh, Srivastava2023-nx, Garcia2024-nq} that enabled us to effectively access the inviscid limit.
Our theoretical analysis complements these simulations, confirming the anomaly and tracing its origin to the nonequilibrium long-range correlations inherent in such systems.

We expect our findings to be general.  
Indeed, our linear analysis immediately predicts anomalous dissipation in other gradient-driven NESSs that exhibit nonequilibrium long-range correlations, ranging from sheared fluids~\cite{Lutsko1985-zb, Nakano2022-kv} to driven electrolyte solutions~\cite{Peraud2016-fd, Peraud2017-xt}.  
This observation leads us to conjecture that  
\textit{anomalous dissipation is a universal feature of gradient-driven nonequilibrium steady states exhibiting long-range nonequilibrium correlations.}
Further developing the analysis of nonlinear fluctuating hydrodynamics and testing this conjecture across diverse systems will be a promising avenue for future work.

Finally, we note that a central question in turbulence research is how the dissipation anomaly is connected to the formation of singular spatial structures.  
This issue has been extensively explored, particularly in the mathematical context of constructing dissipative weak solutions~\cite{De-Lellis2014-pw, Constantin1994-qf, Isett2018-nh}.
Our work extends this fundamental question from the deterministic setting of turbulence to the realm of gradient-driven NESSs described by fluctuating hydrodynamics.
While our finding of anomalous dissipation strongly suggests the presence of underlying singular structures, their precise characterization remains a crucial open problem for future investigation.  
A deeper understanding of these elusive structures would be a key step toward formulating a notion of dissipative weak solutions for stochastic systems.  
We anticipate that the present work opens a fertile direction at the interface of physics and mathematics.

\section*{Acknowledgments}
HN is supported by JSPS KAKENHI Grant No.~JP22K13978.
YM is supported by JSPS KAKENHI Grant No.~JP25K07148 and the Ogawa science and technology foundation.
The numerical computation in this study has been done using the facilities of the Supercomputer Center, the Institute for Solid State Physics, the University of Tokyo. 

\noindent
{\it Data availability.} - All data and codes that support the findings of this study are available in [] for public access.

\bibliographystyle{apsrev4-1}
\bibliography{
    ref_turbu.bib,
    ref_fhd.bib,
    ref_lrc.bib,
    ref_recent.bib,
    ref_others.bib,
    ref_garcia.bib,
    ref_mct.bib
}

@BOOK{Landau1987-eu,
  title     = "Fluid Mechanics",
  author    = "Landau, L D and Lifshitz, E M",
  publisher = "Elsevier",
  year      =  1987,
  url       = "https://books.google.co.jp/books?hl=ja&lr=lang_ja|lang_en&id=eVKbCgAAQBAJ&oi=fnd&pg=PP1&dq=landau+and+lifshitz+fluid+mechanics&ots=YUo0E5MGMG&sig=UUkOSlpkaLzGmU8jFduRuJO9Yo8"
}

@BOOK{De_Zarate2006-xw,
  title     = "Hydrodynamic Fluctuations in Fluids and Fluid Mixtures",
  author    = "de Zarate, Jose M Ortiz and Sengers, Jan V",
  publisher = "Elsevier",
  month     =  apr,
  year      =  2006,
  url       = "https://link.springer.com/article/10.1007/s10765-008-0517-7",
  isbn      =  9780080459431
}

@BOOK{Das2011-ao,
  title     = "Statistical physics of liquids at freezing and beyond",
  author    = "Das, Shankar Prasad",
  publisher = "Cambridge University Press",
  month     =  jun,
  year      =  2011,
  url       = "https://www.cambridge.org/jp/universitypress/subjects/physics/condensed-matter-physics-nanoscience-and-mesoscopic-physics/statistical-physics-liquids-freezing-and-beyond?format=HB&isbn=9780521858397"
}

@ARTICLE{BalboaUsabiaga2012-sh,
  title     = "Staggered Schemes for Fluctuating Hydrodynamics",
  author    = "BalboaUsabiaga, Florencio and Bell, John B and
               Delgado-Buscalioni, Rafael and Donev, Aleksandar and Fai, Thomas
               G and Griffith, Boyce E and Peskin, Charles S",
  journal   = "Multiscale modeling \& simulation",
  publisher = "Society for Industrial and Applied Mathematics",
  volume    =  10,
  number    =  4,
  pages     = "1369--1408",
  month     =  jan,
  year      =  2012,
  url       = "https://doi.org/10.1137/120864520",
  doi       = "10.1137/120864520",
  issn      = "1540-3459"
}

@ARTICLE{Delong2013-fh,
  title     = "Temporal integrators for fluctuating hydrodynamics",
  author    = "Delong, Steven and Griffith, Boyce E and Vanden-Eijnden, Eric and
               Donev, Aleksandar",
  journal   = "Physical Review E",
  publisher = "American Physical Society",
  volume    =  87,
  number    =  3,
  pages     =  033302,
  month     =  mar,
  year      =  2013,
  url       = "https://link.aps.org/doi/10.1103/PhysRevE.87.033302",
  doi       = "10.1103/PhysRevE.87.033302"
}

@ARTICLE{Srivastava2023-nx,
  title    = "Staggered scheme for the compressible fluctuating hydrodynamics of
              multispecies fluid mixtures",
  author   = "Srivastava, Ishan and Ladiges, Daniel R and Nonaka, Andy J and
              Garcia, Alejandro L and Bell, John B",
  journal  = "Physical review. E",
  volume   =  107,
  number   = "1-2",
  pages    =  015305,
  month    =  jan,
  year     =  2023,
  url      = "http://dx.doi.org/10.1103/PhysRevE.107.015305",
  doi      = "10.1103/PhysRevE.107.015305",
  pmid     =  36797914,
  issn     = "2470-0053,2470-0045"
}

@ARTICLE{Garcia2024-nq,
  title         = "An introduction to Computational Fluctuating Hydrodynamics",
  author        = "Garcia, Alejandro L and Bell, John B and Nonaka, Andrew and
                   Srivastava, Ishan and Ladiges, Daniel and Kim, Changho",
  journal = "arXiv:2406.12157",
  year    = 2025,
  url     = "http://arxiv.org/abs/2406.12157"
}

@ARTICLE{Donev2014-jy,
  title     = "Low Mach number fluctuating hydrodynamics of diffusively mixing
               fluids",
  author    = "Donev, Aleksandar and Nonaka, Andy and Sun, Yifei and Fai, Thomas
               and Garcia, Alejandro and Bell, John",
  journal   = "Communications in applied mathematics and computational science",
  publisher = "Mathematical Sciences Publishers",
  volume    =  9,
  number    =  1,
  pages     = "47--105",
  month     =  may,
  year      =  2014,
  url       = "http://dx.doi.org/10.2140/camcos.2014.9.47",
  doi       = "10.2140/camcos.2014.9.47",
  issn      = "1559-3940,2157-5452"
}

@ARTICLE{Donev2015-tm,
  title     = "Low Mach number fluctuating hydrodynamics of multispecies liquid
               mixtures",
  author    = "Donev, Aleksandar and Nonaka, Andy and Bhattacharjee, Amit Kumar
               and Garcia, Alejandro L and Bell, John B",
  journal   = "Physics of fluids",
  publisher = "AIP Publishing",
  volume    =  27,
  number    =  3,
  pages     =  037103,
  month     =  mar,
  year      =  2015,
  url       = "http://dx.doi.org/10.1063/1.4913571",
  doi       = "10.1063/1.4913571",
  issn      = "1070-6631,1089-7666"
}

@ARTICLE{Dorfman1994-cl,
  title     = "Generic Long-Range Correlations in Molecular Fluids",
  author    = "Dorfman, J R and Kirkpatrick, T R and Sengers, J V",
  journal   = "Annual review of physical chemistry",
  publisher = "Annual Reviews",
  volume    =  45,
  number    =  1,
  pages     = "213--239",
  month     =  oct,
  year      =  1994,
  url       = "https://doi.org/10.1146/annurev.pc.45.100194.001241",
  doi       = "10.1146/annurev.pc.45.100194.001241",
  issn      = "0066-426X"
}

@BOOK{Bedeaux2015-lu,
  title     = "Experimental Thermodynamics Volume {X}: Non-equilibrium
               Thermodynamics with Applications",
  author    = "Bedeaux, Dick and Kjelstrup, Signe and Sengers, Jan",
  publisher = "Royal Society of Chemistry",
  month     =  oct,
  year      =  2015,
  url       = "https://books.google.co.jp/books?hl=en&lr=&id=9mooDwAAQBAJ&oi=fnd&pg=PP1&dq=info:tUgPTvVf_4UJ:scholar.google.com&ots=v4qx7pXddD&sig=RihrHvgy97ApUWSqU8naDzmnDDs&redir_esc=y#v=onepage&q&f=false"
}

@ARTICLE{Sengers2024-gz,
  title     = "Mass and thermodiffusion in non-equilibrium fluctuating
               hydrodynamics",
  author    = "Sengers, Jan V",
  journal   = "International journal of thermophysics",
  publisher = "Springer Science and Business Media LLC",
  volume    =  45,
  number    =  9,
  pages     =  132,
  month     =  sep,
  year      =  2024,
  url       = "https://link.springer.com/article/10.1007/s10765-024-03424-1",
  doi       = "10.1007/s10765-024-03424-1",
  issn      = "0195-928X,1572-9567"
}

@ARTICLE{Ronis1982-uk,
  title     = "Nonlinear resonant coupling between shear and heat fluctuations
               in fluids far from equilibrium",
  author    = "Ronis, David and Procaccia, Itamar",
  journal   = "Physical review A",
  publisher = "American Physical Society",
  volume    =  26,
  number    =  3,
  pages     = "1812--1815",
  month     =  sep,
  year      =  1982,
  url       = "https://link.aps.org/doi/10.1103/PhysRevA.26.1812",
  doi       = "10.1103/PhysRevA.26.1812",
  issn      = "1050-2947"
}

@ARTICLE{Lutsko1985-zb,
  title     = "Hydrodynamic fluctuations at large shear rate",
  author    = "Lutsko, J and Dufty, J W",
  journal   = "Physical review A",
  publisher = "APS",
  volume    =  32,
  number    =  5,
  pages     = "3040--3054",
  month     =  nov,
  year      =  1985,
  url       = "http://dx.doi.org/10.1103/physreva.32.3040",
  doi       = "10.1103/physreva.32.3040",
  pmid      =  9896446,
  issn      = "0556-2791"
}

@ARTICLE{Brogioli2001-ct,
  title     = "Diffusive mass transfer by nonequilibrium fluctuations: Fick's
               law revisited",
  author    = "Brogioli, D and Vailati, A",
  journal   = "Physical review E",
  publisher = "American Physical Society (APS)",
  volume    =  63,
  number    = "1 Pt 1",
  pages     =  012105,
  month     =  jan,
  year      =  2001,
  url       = "https://journals.aps.org/pre/abstract/10.1103/PhysRevE.63.012105",
  doi       = "10.1103/PhysRevE.63.012105",
  pmid      =  11304296,
  issn      = "1539-3755,1550-2376"
}

@ARTICLE{Donev2011-hf,
  title     = "Diffusive transport by thermal velocity fluctuations",
  author    = "Donev, Aleksandar and Bell, John B and de la Fuente, Anton and
               Garcia, Alejandro L",
  journal   = "Physical review letters",
  publisher = "APS",
  volume    =  106,
  number    =  20,
  pages     =  204501,
  month     =  may,
  year      =  2011,
  url       = "http://dx.doi.org/10.1103/PhysRevLett.106.204501",
  doi       = "10.1103/PhysRevLett.106.204501",
  pmid      =  21668233,
  issn      = "0031-9007,1079-7114"
}

@ARTICLE{Kirkpatrick2015-ai,
  title     = "Nonequilibrium is different",
  author    = "Kirkpatrick, T R and Dorfman, J R",
  journal   = "Physical review E",
  publisher = "American Physical Society (APS)",
  volume    =  92,
  number    =  2,
  pages     =  022109,
  month     =  aug,
  year      =  2015,
  url       = "https://journals.aps.org/pre/abstract/10.1103/PhysRevE.92.022109",
  doi       = "10.1103/PhysRevE.92.022109",
  pmid      =  26382346,
  issn      = "1539-3755,1550-2376"
}

@ARTICLE{Kirkpatrick2021-rv,
  title    = "Rigidity and Superfast Signal Propagation in Fluids and Solids in
              Non-Equilibrium Steady States",
  author   = "Kirkpatrick, T R and Belitz, D and Dorfman, J R",
  journal  = "The journal of physical chemistry B",
  volume   =  125,
  number   =  27,
  pages    = "7499--7507",
  month    =  jul,
  year     =  2021,
  url      = "http://dx.doi.org/10.1021/acs.jpcb.0c11283",
  doi      = "10.1021/acs.jpcb.0c11283",
  pmid     =  34191519,
  issn     = "1520-6106,1520-5207"
}

@article{Peraud2016-fd,
  title     = {Low Mach number fluctuating hydrodynamics for electrolytes},
  author    = {P\'eraud, Jean-Philippe and Nonaka, Andy and Chaudhri, Anuj and Bell, John B. and Donev, Aleksandar and Garcia, Alejandro L.},
  journal   = {Physical review fluids},
  volume    = {1},
  issue     = {7},
  pages     = {074103},
  numpages  = {27},
  year      = {2016},
  month     = {Nov},
  publisher = {American Physical Society},
  doi       = {10.1103/PhysRevFluids.1.074103},
  url       = {https://link.aps.org/doi/10.1103/PhysRevFluids.1.074103}
}

@ARTICLE{Peraud2017-xt,
  title     = "Fluctuation-enhanced electric conductivity in electrolyte
               solutions",
  author    = "Péraud, Jean-Philippe and Nonaka, Andrew J and Bell, John B and
               Donev, Aleksandar and Garcia, Alejandro L",
  journal   = "Proceedings of the national academy of sciences of the United
               States of America",
  publisher = "National Acad Sciences",
  volume    =  114,
  number    =  41,
  pages     = "10829--10833",
  month     =  oct,
  year      =  2017,
  url       = "http://dx.doi.org/10.1073/pnas.1714464114",
  doi       = "10.1073/pnas.1714464114",
  pmc       = "PMC5642729",
  pmid      =  28973890,
  issn      = "0027-8424,1091-6490"
}

@ARTICLE{Nakano2022-kv,
  title     = "Molecular dynamics study of shear-induced long-range correlations
               in simple fluids",
  author    = "Nakano, Hiroyoshi and Minami, Yuki",
  journal   = "Physical review research",
  publisher = "American Physical Society",
  volume    =  4,
  number    =  2,
  pages     =  023147,
  month     =  may,
  year      =  2022,
  url       = "https://link.aps.org/doi/10.1103/PhysRevResearch.4.023147",
  doi       = "10.1103/PhysRevResearch.4.023147"
}

@ARTICLE{Nakano2025-zw,
  title   = "Long-range correlations under temperature gradients: A molecular
             dynamics study of simple fluids",
  author  = "Nakano, Hiroyoshi and Yokota, Kazuma",
  journal = "Physical review E",
  publisher = "American Physical Society",
  volume  =  111,
  number  =  6,
  pages   =  "L063401",
  month   =  jun,
  year    =  2025,
  url     = "http://dx.doi.org/10.1103/PhysRevE.111.L063401",
  doi     = "10.1103/PhysRevE.111.L063401"
}

@ARTICLE{Bussoletti2025-qh,
  title     = "Emergence of long-range non-equilibrium correlations in free
               liquid diffusion",
  author    = "Bussoletti, Marco and Gallo, Mirko and Jafari, Amir and Eyink,
               Gregory L",
  journal   = "The Journal of chemical physics",
  publisher = "AIP Publishing",
  volume    =  163,
  number    =  16,
  pages     =  164509,
  month     =  oct,
  year      =  2025,
  url       = "http://dx.doi.org/10.1063/5.0292952",
  doi       = "10.1063/5.0292952",
  pmid      =  41143499,
  issn      = "0021-9606,1089-7690"
}

@ARTICLE{Bussoletti2025-jd,
  title   = "Non-gaussian statistics of concentration fluctuations in free liquid diffusion",
  author  = "Bussoletti, Marco and Gallo, Mirko and Jafari, Amir and
             Eyink, Gregory L",
  journal = "arXiv:2509.25511",
  year    = 2025,
  url     = "http://arxiv.org/abs/2509.25511"
}

@ARTICLE{Vailati1997-ez,
  title     = "Giant fluctuations in a free diffusion process",
  author    = "Vailati, A and Giglio, M",
  journal   = "Nature",
  publisher = "nature.com",
  volume    = 390,
  pages     = "262–265",
  year      =  1997,
  url       = "https://www.nature.com/articles/36803",
  issn      = "0028-0836"
}

@ARTICLE{Takacs2011-gg,
  title     = "Thermal Fluctuations in a Layer of Liquid {CS}$_{2}$ Subjected to
               Temperature Gradients with and without the Influence of Gravity",
  author    = "Takacs, Christopher J and Vailati, Alberto and Cerbino, Roberto
               and Mazzoni, Stefano and Giglio, Marzio and Cannell, David S",
  journal   = "Physical review letters",
  publisher = "American Physical Society",
  volume    =  106,
  number    =  24,
  pages     =  244502,
  month     =  jun,
  year      =  2011,
  url       = "https://link.aps.org/doi/10.1103/PhysRevLett.106.244502",
  doi       = "10.1103/PhysRevLett.106.244502",
  issn      = "0031-9007"
}

@ARTICLE{Vailati2011-rz,
  title     = "Fractal fronts of diffusion in microgravity",
  author    = "Vailati, Alberto and Cerbino, Roberto and Mazzoni, Stefano and
               Takacs, Christopher J and Cannell, David S and Giglio, Marzio",
  journal   = "Nature communications",
  publisher = "Springer Science and Business Media LLC",
  volume    =  2,
  number    =  1,
  pages     =  290,
  year      =  2011,
  url       = "https://www.nature.com/articles/ncomms1290",
  doi       = "10.1038/ncomms1290",
  pmc       = "PMC3220270",
  pmid      =  21505446,
  issn      = "2041-1723,2041-1723"
}

@ARTICLE{Vailati2012-yz,
  title     = "Gradient-driven fluctuations in microgravity",
  author    = "Vailati, A and Cerbino, R and Mazzoni, S and Giglio, M and
               Takacs, C J and Cannell, D S",
  journal   = "Journal of physics: condensed matter",
  publisher = "iopscience.iop.org",
  volume    =  24,
  number    =  28,
  pages     =  284134,
  month     =  jul,
  year      =  2012,
  url       = "http://dx.doi.org/10.1088/0953-8984/24/28/284134",
  doi       = "10.1088/0953-8984/24/28/284134",
  pmid      =  22739247,
  issn      = "0953-8984,1361-648X"
}

@ARTICLE{Wada2003-je,
  title     = "Anomalous pressure in fluctuating shear flow",
  author    = "Wada, Hirofumi and Sasa, Shin-Ichi",
  journal   = "Physical review E",
  publisher = "APS",
  volume    =  67,
  number    = "6",
  pages     =  065302,
  month     =  jun,
  year      =  2003,
  url       = "http://dx.doi.org/10.1103/PhysRevE.67.065302",
  doi       = "10.1103/PhysRevE.67.065302",
  pmid      =  16241294,
  issn      = "1539-3755"
}

@ARTICLE{Kirkpatrick2013-ic,
  title     = "Giant Casimir effect in fluids in nonequilibrium steady states",
  author    = "Kirkpatrick, T R and Ortiz de Zárate, J M and Sengers, J V",
  journal   = "Physical review letters",
  publisher = "APS",
  volume    =  110,
  number    =  23,
  pages     =  235902,
  month     =  jun,
  year      =  2013,
  url       = "http://dx.doi.org/10.1103/PhysRevLett.110.235902",
  doi       = "10.1103/PhysRevLett.110.235902",
  pmid      =  25167514,
  issn      = "0031-9007,1079-7114"
}

@ARTICLE{Kirkpatrick2015-rg,
  title     = "Nonequilibrium Casimir-like Forces in Liquid Mixtures",
  author    = "Kirkpatrick, T R and Ortiz de Zárate, J M and Sengers, J V",
  journal   = "Physical review letters",
  publisher = "APS",
  volume    =  115,
  number    =  3,
  pages     =  035901,
  month     =  jul,
  year      =  2015,
  url       = "http://dx.doi.org/10.1103/PhysRevLett.115.035901",
  doi       = "10.1103/PhysRevLett.115.035901",
  pmid      =  26230803,
  issn      = "0031-9007,1079-7114"
}

@ARTICLE{Kirkpatrick2016-ev,
  title    = "Nonequilibrium fluctuation-induced Casimir pressures in liquid
              mixtures",
  author   = "Kirkpatrick, T R and Ortiz de Zárate, J M and Sengers, J V",
  journal  = "Physical review E",
  volume   =  93,
  number   =  3,
  pages    =  032117,
  month    =  mar,
  year     =  2016,
  url      = "http://dx.doi.org/10.1103/PhysRevE.93.032117",
  doi      = "10.1103/PhysRevE.93.032117",
  pmid     =  27078302,
  issn     = "2470-0053,2470-0045"
}

@ARTICLE{Croccolo2016-mn,
  title     = "Non-local fluctuation phenomena in liquids",
  author    = "Croccolo, F and Ortiz de Zárate, J M and Sengers, J V",
  journal   = "The European physical journal. E",
  publisher = "Springer",
  volume    =  39,
  number    =  12,
  pages     =  125,
  month     =  dec,
  year      =  2016,
  url       = "http://dx.doi.org/10.1140/epje/i2016-16125-3",
  doi       = "10.1140/epje/i2016-16125-3",
  pmid      =  27987100,
  issn      = "1292-8941,1292-895X"
}

@ARTICLE{Kawasaki1970-vo,
  title     = "Long time behavior of the velocity autocorrelation function",
  author    = "Kawasaki, K",
  journal   = "Physics letters A",
  publisher = "Elsevier",
  volume    =  32,
  pages     = "379--380",
  month     =  aug,
  year      =  1970,
  url       = "http://dx.doi.org/10.1016/0375-9601(70)90009-5",
  doi       = "10.1016/0375-9601(70)90009-5"
}

@ARTICLE{Kawasaki1971-am,
  title     = "Application of extended mode-coupling theory to long-time
               behavior of correlation functions",
  author    = "Kawasaki, K",
  journal   = "Progress of theoretical physics",
  publisher = "academic.oup.com",
  volume    =  45,
  pages     = "1691--1692",
  month     =  may,
  year      =  1971,
  url       = "http://dx.doi.org/10.1143/PTP.46.1299",
  doi       = "10.1143/PTP.46.1299"
}

@ARTICLE{Bedeaux1974-da,
  title     = "Renormalization of the diffusion coefficient in a fluctuating
               fluid {I}",
  author    = "Bedeaux, D and Mazur, P",
  journal   = "Physica",
  publisher = "North-Holland",
  volume    =  73,
  number    =  3,
  pages     = "431--458",
  month     =  may,
  year      =  1974,
  url       = "https://www.sciencedirect.com/science/article/abs/pii/0031891474901049?via%3Dihub",
  doi       = "10.1016/0031-8914(74)90104-9",
  issn      = "0031-8914"
}

@ARTICLE{Mazur1974-bs,
  title     = "Renormalization of the diffusion coefficient in a fluctuating
               fluid {II}",
  author    = "Mazur, P and Bedeaux, D",
  journal   = "Physica",
  publisher = "Elsevier BV",
  volume    =  75,
  number    =  1,
  pages     = "79--99",
  month     =  jul,
  year      =  1974,
  url       = "https://www.sciencedirect.com/science/article/abs/pii/0031891474902936?via%3Dihub",
  doi       = "10.1016/0031-8914(74)90293-6",
  issn      = "0031-8914,1873-1767",
  language  = "en"
}

@ARTICLE{Pomeau1975-ll,
  title     = "Time dependent correlation functions and mode-mode coupling
               theories",
  author    = "Pomeau, Y and Résibois, P",
  journal   = "Physics reports",
  publisher = "Elsevier",
  volume    =  19,
  number    =  2,
  pages     = "63--139",
  month     =  jun,
  year      =  1975,
  url       = "https://www.sciencedirect.com/science/article/pii/0370157375900198",
  doi       = "10.1016/0370-1573(75)90019-8",
  issn      = "0370-1573"
}

@ARTICLE{Forster1977-lr,
  title     = "Large-distance and long-time properties of a randomly stirred
               fluid",
  author    = "Forster, Dieter and Nelson, David R and Stephen, Michael J",
  journal   = "Physical review. A",
  publisher = "American Physical Society",
  volume    =  16,
  number    =  2,
  pages     = "732--749",
  month     =  aug,
  year      =  1977,
  url       = "https://link.aps.org/doi/10.1103/PhysRevA.16.732",
  doi       = "10.1103/PhysRevA.16.732",
  issn      = "1050-2947"
}

@BOOK{degroot2013-ib,
  title     = "Non-equilibrium thermodynamics",
  author    = "De Groot, S. R. and Mazur, P.",
  publisher = "Courier Corporation",
  year      =  2013,
  url       = "https://books.google.co.jp/books?hl=ja&lr=lang_ja|lang_en&id=mfFyG9jfaMYC&oi=fnd&pg=PP1&dq=de+groot+mazur&ots=ii0iuvE-qu&sig=9j8Ya4zjx4c6U_z6x0WVmZFMMHw"
}

@BOOK{chaikin1995-pr,
  title      = "Principles of Condensed Matter Physics",
  author     = "Chaikin, P. M. and Lubensky, T. C.",
  publisher  = "Cambridge University Press",
  year       = 1995,
  url        = "https://www.cambridge.org/core/books/principles-of-condensed-matter-physics/70C3D677A9B5BEC4A77CBBD0A8A23E64"
}

@BOOK{Mazenko2006,
  title     = "Nonequilibrium Statistical Mechanics",
  author    = "Mazenko, Gene F.",
  publisher = "Wiley-VCH",
  year      = "2006",
  doi       = "10.1002/9783527618958",
  url       = "https://onlinelibrary.wiley.com/doi/book/10.1002/9783527618958"
}

@ARTICLE{McMullen2022-xb,
  title     = "Navier-Stokes Equations Do Not Describe the Smallest Scales of
               Turbulence in Gases",
  author    = "McMullen, Ryan M and Krygier, Michael C and Torczynski, John R
               and Gallis, Michael A",
  journal   = "Physical review letters",
  publisher = "APS",
  volume    =  128,
  number    =  11,
  pages     =  114501,
  month     =  mar,
  year      =  2022,
  url       = "http://dx.doi.org/10.1103/PhysRevLett.128.114501",
  doi       = "10.1103/PhysRevLett.128.114501",
  pmid      =  35363027,
  issn      = "0031-9007,1079-7114"
}

@ARTICLE{Bell2022-en,
  title     = "Thermal fluctuations in the dissipation range of homogeneous
               isotropic turbulence",
  author    = "Bell, J B and Nonaka, A and Garcia, A L and Eyink, G",
  journal   = "Journal of fluid mechanics",
  publisher = "cambridge.org",
  volume    =  939,
  pages     =  "A12",
  year      =  2022,
  url       = "https://www.cambridge.org/core/journals/journal-of-fluid-mechanics/article/thermal-fluctuations-in-the-dissipation-range-of-homogeneous-isotropic-turbulence/2F84D45D9339AFB84A56D6CC55DC1277",
  issn      = "0022-1120"
}

@ARTICLE{Eyink2022-fc,
  title     = "High Schmidt-number turbulent advection and giant concentration
               fluctuations",
  author    = "Eyink, Gregory and Jafari, Amir",
  journal   = "Physical review research",
  publisher = "American Physical Society (APS)",
  volume    =  4,
  number    =  2,
  pages     =  023246,
  month     =  jun,
  year      =  2022,
  url       = "http://dx.doi.org/10.1103/physrevresearch.4.023246",
  doi       = "10.1103/physrevresearch.4.023246",
  issn      = "2643-1564",
  language  = "en"
}

@ARTICLE{Bandak2022-rk,
  title    = "Dissipation-range fluid turbulence and thermal noise",
  author   = "Bandak, Dmytro and Goldenfeld, Nigel and Mailybaev, Alexei A and
              Eyink, Gregory",
  journal  = "Physical review E",
  volume   =  105,
  number   = "6-2",
  pages    =  065113,
  month    =  jun,
  year     =  2022,
  url      = "http://dx.doi.org/10.1103/PhysRevE.105.065113",
  doi      = "10.1103/PhysRevE.105.065113",
  pmid     =  35854607,
  issn     = "2470-0053,2470-0045"
}

@ARTICLE{Bandak2024-al,
  title     = "Spontaneous stochasticity amplifies even thermal noise to the
               largest scales of turbulence in a few eddy turnover times",
  author    = "Bandak, Dmytro and Mailybaev, Alexei A and Eyink, Gregory L and
               Goldenfeld, Nigel",
  journal   = "Physical review letters",
  publisher = "American Physical Society (APS)",
  volume    =  132,
  number    =  10,
  pages     =  104002,
  month     =  mar,
  year      =  2024,
  url       = "https://journals.aps.org/prl/abstract/10.1103/PhysRevLett.132.104002",
  doi       = "10.1103/PhysRevLett.132.104002",
  pmid      =  38518311,
  issn      = "0031-9007,1079-7114"
}

@ARTICLE{Ishan2025-fw,
  title         = "Molecular Fluctuations Inhibit Intermittency in Compressible
                   Turbulence",
  author  = "Ishan, Srivastava and Andrew, J Nonaka and Weiqun, Zhang and
                   Alejandro, L Garcia and John, B Bell",
  journal = "arXiv:2501.06396",
  year    = 2025,
  url     = "http://arxiv.org/abs/2501.06396"
}

@ARTICLE{Rodriguez-Fernandez2022-qw,
  title    = "Anomalous ballistic scaling in the tensionless or inviscid
              Kardar-Parisi-Zhang equation",
  author   = "Rodríguez-Fernández, Enrique and Santalla, Silvia N and Castro,
              Mario and Cuerno, Rodolfo",
  journal  = "Physical review E",
  volume   =  106,
  number   = "2-1",
  pages    =  024802,
  month    =  aug,
  year     =  2022,
  url      = "http://dx.doi.org/10.1103/physreve.106.024802",
  doi      = "10.1103/PhysRevE.106.024802",
  pmid     =  36109999,
  issn     = "2470-0053,2470-0045"
}

@ARTICLE{Cartes2022-ur,
  title     = "The Galerkin-truncated Burgers equation: crossover from
               inviscid-thermalized to Kardar-Parisi-Zhang scaling",
  author    = "Cartes, C and Tirapegui, E and Pandit, R and Brachet, M",
  journal   = "Philosophical transactions of the royal society A",
  publisher = "The Royal Society",
  volume    =  380,
  number    =  2219,
  pages     =  20210090,
  month     =  mar,
  year      =  2022,
  url       = "http://dx.doi.org/10.1098/rsta.2021.0090",
  doi       = "10.1098/rsta.2021.0090",
  pmid      =  35094560,
  issn      = "1364-503X,1471-2962"
}

@ARTICLE{Fontaine2023-nr,
  title    = "Unpredicted Scaling of the One-Dimensional Kardar-Parisi-Zhang
              Equation",
  author   = "Fontaine, Côme and Vercesi, Francesco and Brachet, Marc and Canet,
              Léonie",
  journal  = "Physical review letters",
  volume   =  131,
  number   =  24,
  pages    =  247101,
  month    =  dec,
  year     =  2023,
  url      = "http://dx.doi.org/10.1103/PhysRevLett.131.247101",
  doi      = "10.1103/PhysRevLett.131.247101",
  pmid     =  38181147,
  issn     = "0031-9007,1079-7114"
}

@ARTICLE{Gosteva2025-cq,
  title   = "Emergent dynamical scaling in the inviscid limit of {3D}
                   stochastic Navier-Stokes equation with thermal noise",
  author  = "Gosteva, Liubov and Brachet, Marc and Canet, Léonie",
  journal = "arXiv:2507.05811",
  year    = 2025,
  url     = "http://arxiv.org/abs/2507.05811"
}

@ARTICLE{Nakano2025-tj,
  title   = "Looking at bare transport coefficients in fluctuating
                   hydrodynamics",
  author  = "Nakano, Hiroyoshi and Minami, Yuki and Saito, Keiji",
  journal = "arXiv:2502.15241",
  year    = 2025,
  url     = "http://arxiv.org/abs/2502.15241"
}

@BOOK{Frisch1995-rs,
  title     = "Turbulence: the legacy of A. N. Kolmogorov",
  author    = "Frisch, U and Kolmogorov, A N",
  publisher = "Cambridge university press",
  year      =  1995,
  url       = "hhttps://www.cambridge.org/highereducation/books/turbulence/FD8C5E35E5F1CA850E017461942A59AC#overview"
}

@ARTICLE{Kolmogorov1941-gz,
  title   = "Dissipation of energy in the locally isotropic turbulence",
  author  = "Kolmogorov, A",
  journal = "Proceedings of the royal society of London",
  volume  =  434,
  pages   = "15--17",
  month   =  apr,
  year    =  1941,
  url     = "http://dx.doi.org/10.1098/rspa.1991.0076",
  doi     = "10.1098/rspa.1991.0076",
  issn    = "0080-4630,0002-3264"
}

@ARTICLE{Kolmogorov1941-kx,
  title   = "The local structure of turbulence in incompressible viscous fluid
             for very large Reynolds' numbers",
  author  = "Kolmogorov, A",
  journal = "Doklady Akademii nauk SSSR",
  volume  =  30,
  pages   = "301--305",
  year    =  1941,
  url     = "https://ui.adsabs.harvard.edu/abs/1941DoSSR..30..301K/abstract",
  issn    = "0002-3264"
}

@ARTICLE{Onsager1949-ze,
  title     = "Statistical hydrodynamics",
  author    = "Onsager, L",
  journal   = "Il Nuovo cimento",
  publisher = "Springer Science and Business Media LLC",
  volume    =  6,
  number    = "S2",
  pages     = "279--287",
  month     =  mar,
  year      =  1949,
  url       = "http://dx.doi.org/10.1007/bf02780991",
  doi       = "10.1007/bf02780991",
  issn      = "0029-6341,1827-6121"
}

@INCOLLECTION{De-Lellis2014-pw,
  title     = "Continuous dissipative Euler flows and a conjecture of Onsager",
  author    = "De Lellis, C and Székelyhidi, L",
  booktitle = "European congress of mathematics",
  publisher = "European Mathematical Society",
  address   = "Zürich, Switzerland",
  edition   =  1,
  month     =  jan,
  year      =  2014,
  url       = "http://dx.doi.org/10.4171/120",
  doi       = "10.4171/120",
  isbn      =  "9783037191200"
}

@ARTICLE{Constantin1994-qf,
  title     = "Onsager's conjecture on the energy conservation for solutions of
               Euler's equation",
  author    = "Constantin, Peter and Weinan, E and Titi, Edriss S",
  journal   = "Communications in mathematical physics",
  publisher = "Springer Science and Business Media LLC",
  volume    =  165,
  number    =  1,
  pages     = "207--209",
  month     =  oct,
  year      =  1994,
  url       = "http://dx.doi.org/10.1007/bf02099744",
  doi       = "10.1007/bf02099744",
  issn      = "0010-3616,1432-0916"
}

@ARTICLE{Isett2018-nh,
  title     = "A proof of Onsager's conjecture",
  author    = "Isett, Philip",
  journal   = "Annals of mathematics",
  publisher = "Annals of Mathematics",
  volume    =  188,
  number    =  3,
  pages     =  871,
  month     =  nov,
  year      =  2018,
  url       = "http://dx.doi.org/10.4007/annals.2018.188.3.4",
  doi       = "10.4007/annals.2018.188.3.4",
  issn      = "0003-486X,1939-8980"
}

@ARTICLE{Sreenivasan1998-sb,
  title     = "An update on the energy dissipation rate in isotropic turbulence",
  author    = "Sreenivasan, K",
  journal   = "Physics of fluids",
  publisher = "pubs.aip.org",
  volume    = 10,
  pages     = 528,
  month     =  jun,
  year      =  1998,
  url       = "http://dx.doi.org/10.1063/1.869575",
  doi       = "10.1063/1.869575",
  issn      = "0967-0653,1878-6731"
}

@ARTICLE{Kaneda2003-ni,
  title     = "Energy dissipation rate and energy spectrum in high resolution
               direct numerical simulations of turbulence in a periodic box",
  author    = "Kaneda, Y and Ishihara, T and Yokokawa, M and Itakura, K and Uno,
               Atsuya",
  journal   = "Physics of fluids",
  publisher = "pubs.aip.org",
  volume    =  15,
  pages     =  "L21",
  month     =  jan,
  year      =  2003,
  url       = "http://dx.doi.org/10.1063/1.1539855",
  doi       = "10.1063/1.1539855"
}

@ARTICLE{Yeung2012-fo,
  title     = "Dissipation, enstrophy and pressure statistics in turbulence
               simulations at high Reynolds numbers",
  author    = "Yeung, P and Donzis, D and Sreenivasan, K",
  journal   = "Journal of fluid mechanics",
  publisher = "cambridge.org",
  volume    =  700,
  pages     = "5--15",
  month     =  feb,
  year      =  2012,
  url       = "http://dx.doi.org/10.1017/jfm.2012.5",
  doi       = "10.1017/jfm.2012.5",
  issn      = "0022-1120,1469-7645"
}

@ARTICLE{Iyer2025-kb,
  title     = "Turbulence without walls: Whither the zeroth law of turbulence?",
  author    = "Iyer, Kartik P and Drivas, Theodore D and Eyink, Gregory L and
               Sreenivasan, Katepalli R",
  journal   = "Physical review letters",
  publisher = "American Physical Society (APS)",
  volume    =  135,
  number    =  13,
  pages     =  134001,
  month     =  sep,
  year      =  2025,
  url       = "http://dx.doi.org/10.1103/xpwj-txlp",
  doi       = "10.1103/xpwj-txlp",
  pmid      =  41076680,
  issn      = "0031-9007,1079-7114"
}

\clearpage
\begin{center}
{\large \bf End Matter}
\end{center}

\section{Details of the numerical simulation}
The numerical simulations used in the main text were performed using the following scheme.

\sectionprl{low-Mach-Number regime}
To efficiently simulate the incompressible dynamics discussed in the main text, we employ a compressible fluctuating hydrodynamics framework in the low-Mach-number regime, realized by choosing a sufficiently large sound speed $c_T$.
This approach is often computationally more direct, as it avoids the numerical challenge of enforcing the incompressibility condition ($\nabla \cdot \bm{v} = 0$) at each time step.
As validated in the SM~\footnotemark[1], the effects of fluid compressibility (longitudinal modes) on the dissipation rates vanish in the limit of infinite sound speed, $c_T \to \infty$.
Our approach therefore provides a faithful numerical realization of the incompressible dynamics targeted in the main text.

In our simulations, the velocity field $\bm{v}$ is generated by the compressible fluctuating Navier--Stokes equations:
\begin{align}
    & \frac{\partial \rho}{\partial t} + \nabla \cdot (\rho \bm{v}) = 0, \label{eq:app_sto_cont} \\[2pt]
    & \rho \frac{\partial \bm{v}}{\partial t} = - \nabla p + \eta_0 \nabla^2 \bm{v} \nonumber \\
    & \hspace{1.5cm} + [\zeta_0 + (1-\tfrac{2}{d})\eta_0]\nabla(\nabla \cdot \bm{v}) - \nabla \cdot \bm{\Pi}_{\rm{ran}}. \label{eq:app_sto_ns}
\end{align}
Here, $\eta_0$ and $\zeta_0$ denote the shear and bulk viscosities, respectively.
We adopt a barotropic equation of state $p = c_T^2 \rho$, where the isothermal sound speed $c_T$ is set to a large value to minimize compressibility effects.
Using this velocity field, we solve the advection-diffusion equation for the scalar fluctuation $\delta\psi$:
\begin{align}
    \frac{\partial (\delta\psi)}{\partial t} + G v_y + \bm{v}\!\cdot\!\nabla (\delta\psi)
    &= D_0 \nabla^2 (\delta\psi) - \nabla\!\cdot\!\bm{J}_{\mathrm{ran}}. \label{eq:app_sto_cd_fluct}
\end{align}
While Eq.~(\ref{eq:app_sto_cd_fluct}) takes the same form as Eq.~(\ref{eq:sto_cd_fluct}) in the main text, the velocity field $\bm{v}$ used here is a solution to the compressible Navier--Stokes equations, unlike the strictly incompressible fluid.

The stochastic stress $\bm{\Pi}_{\mathrm{ran}}$ obeys the fluctuation--dissipation theorem for a compressible fluid, while the definition of $\bm{J}_{\mathrm{ran}}$ remains identical to that in the main text:
\begin{align}
    & \langle \Pi_{\mathrm{ran}}^{ab}(\bm{r},t)\Pi_{\mathrm{ran}}^{cd}(\bm{r}',t') \rangle = 2 k_B T \delta(\bm{r}-\bm{r}')\delta(t-t') \nonumber \\
    & \hspace{0.5cm} \times \Big[\eta_0(\delta^{ac}\delta^{bd}+\delta^{ad}\delta^{bc}) + (\zeta_0 - \tfrac{2}{d}\eta_0)\delta^{ab}\delta^{cd}\Big], \\[2pt]
    & \langle J_{\mathrm{ran}}^{a}(\bm{r},t) J_{\mathrm{ran}}^{b}(\bm{r}',t') \rangle
    = 2 k_B T \chi D_0\, \delta^{ab}\delta(\bm{r}-\bm{r}')\delta(t-t').
\end{align}

\sectionprl{Numerical implementation}
We solve Eqs.~(\ref{eq:app_sto_cont})–(\ref{eq:app_sto_cd_fluct}) using the scheme developed by Garcia, Bell, \textit{et al.}~\cite{BalboaUsabiaga2012-sh, Delong2013-fh, Srivastava2023-nx, Garcia2024-nq}.
The scheme employs a staggered lattice for spatial discretization and a low-storage third-order stochastic Runge-Kutta method for time integration.
A detailed description of our implementation, following the same methodology, is given in Ref.~\cite{Nakano2025-tj}.

We define the fundamental units of our system by setting the density $\rho_0 = 1.0$, the temperature $k_B T = 1.0$, and fixing the system size to $L_x = L_y = 2048$.
The spatial domain is discretized into a $64 \times 64$ lattice, corresponding to a grid spacing of $a_{\rm uv} = 32.0$.
The time step for the temporal discretization is set to $dt = 0.01$.
In addition, we set $c_T^2 = 1000.0$ to enforce incompressibility, and use $\chi=0.01$ to suppress equilibrium fluctuations.
We also set $\zeta_0 = \eta_0$.

%---------------flguire------------------
\begin{figure}[t]
\begin{center}
\includegraphics[scale=1.0]{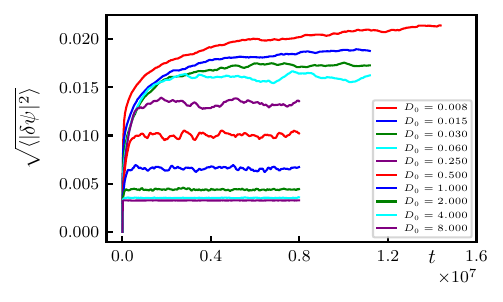}
\end{center}
\vspace{-0.5cm}
\caption{
Relaxation to the non-equilibrium steady state in the numerical simulation.
The time evolution of the mean-squared fluctuation, $\langle |\delta \psi|^2 \rangle$, is plotted as a function of time $t$.
}
\label{supfig1}
\end{figure}
%---------------flguire------------------
%---------------table------------------
\begin{table}[b]
\centering
\begin{tabularx}{0.9\columnwidth}{@{\extracolsep{\fill}}c|c|c}
\hline
$D_0$ & Relaxation time & Number of samples \\
\hline
$\qquad0.008\qquad$ & $12.8\times10^6$ & $216$\\
$\qquad0.015\qquad$ & $9.6\times10^6$ & $216$\\
$\qquad0.030\qquad$ & $9.6\times10^6$ & $216$\\
$\qquad0.060\qquad$ & $9.6\times10^6$ & $216$\\
$\qquad0.120\qquad$ & $9.6\times10^6$ & $216$\\
$\qquad0.250\qquad$ & $6.4\times10^6$ & $144$\\
$\qquad0.500\qquad$ & $6.4\times10^6$ & $144$\\
$\qquad1.000\qquad$ & $6.4\times10^6$ & $144$\\
$\qquad2.000\qquad$ & $6.4\times10^6$ & $144$\\
$\qquad4.000\qquad$ & $6.4\times10^6$ & $144$\\
$\qquad8.000\qquad$ & $6.4\times10^6$ & $144$\\
\hline
\end{tabularx}
\caption
{
Number of time steps to reach a non-equilibrium steady state and number of samples for averaging.
}
\label{emtab1}
\end{table}
%---------------table------------------
\sectionprl{Observation protocol}
Each simulation is initialized with $\rho(\bm{r})=\rho_0$, $\delta\psi(\bm{r})=0$, and $\bm{v}(\bm{r})=\bm{0}$.  
The measurements follow a two-stage protocol.  
First, the system is evolved through a relaxation run until it reaches a nonequilibrium steady state; the duration of this stage is listed in Table~\ref{emtab1}.  
Steady-state attainment is confirmed by monitoring the time evolution of $\langle |\delta\psi|^2\rangle$, as illustrated in Fig.~\ref{supfig1}.

After the relaxation stage, a production run of $1.6\times 10^8$ steps 
($1.6\times 10^6$ time units) is conducted.  
Observables are sampled every $10^5$ steps.  
To improve statistics, the reported results are averaged over multiple independent realizations with different noise seeds.  
The number of realizations used for each parameter set is listed in Table~\ref{emtab1}.

\section{Decomposition of the fluctuations}
In this Appendix, we clarify the definitions of the equilibrium and nonequilibrium components of the scalar fluctuation, $\delta\psi_{\mathrm{eq}}$ and $\delta\psi_{\mathrm{neq}}$.
The equilibrium component $\delta\psi_{\mathrm{eq}}$ is defined as the fluctuation that remains even in the absence of the external gradient ($G=0$),  
and is therefore governed by the equilibrium dynamics of the system:
\begin{align}
    \frac{\partial (\delta\psi_{\rm eq})}{\partial t} + \bm{v} \cdot \nabla (\delta\psi_{\rm eq}) = D_0 \nabla^2 (\delta\psi_{\rm eq}) - \nabla \cdot \bm{J}_{\rm{ran}}.
    \label{eq:appendix_sto_cd_fluct_eq}
\end{align}
The nonequilibrium fluctuation $\delta\psi_{\rm neq}$ is the additional fluctuation induced by the gradient $G$, which is defined as $\delta\psi_{\rm neq} := \delta\psi - \delta\psi_{\rm eq}$.
Since the full equation for $\delta\psi$ [Eq.~(\ref{eq:sto_cd_fluct})] is linear in $\delta\psi$, we obtain the equation for $\delta\psi_{\rm neq}$ by simply subtracting Eq.~(\ref{eq:appendix_sto_cd_fluct_eq}) from Eq.~(\ref{eq:sto_cd_fluct}):
\begin{align}
    \frac{\partial (\delta\psi_{\rm neq})}{\partial t} + v^y G + \bm{v} \cdot \nabla (\delta\psi_{\rm neq}) = D_0 \nabla^2 (\delta\psi_{\rm neq}).
    \label{eq:appendix_neq_fluctuation_eq}
\end{align}
Since the velocity field $\bm{v}$ is a stochastic variable, the solution $\delta\psi_{\rm neq}$ to this equation is also a stochastic variable.

% \newpage

%---------------flguire------------------
\begin{figure}[t]
\begin{center}
\includegraphics[scale=1.00]{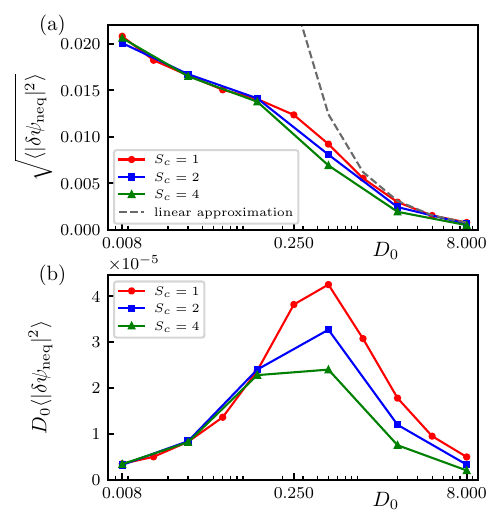}
\end{center}
\vspace{-0.5cm}
\caption{
Numerical results for the amplitude of the nonequilibrium fluctuations.
Simulation parameters are the same as in Fig.~\ref{fig1}.
(a) The root-mean-square (RMS) amplitude, $\sqrt{\langle |\delta \psi_{\rm neq}|^2\rangle}$, as a function of $D_0$ for several Schmidt numbers $S_c$.
(b) The product of the diffusion coefficient and the mean-squared amplitude, $D_0 \langle |\delta \psi_{\rm neq}|^2 \rangle$, as a function of $D_0$.
}
\label{emfig2}
\end{figure}
%---------------flguire------------------
\section{Amplitude of the nonequilibrium fluctuations}
In the main text, we demonstrated that the gradient of the nonequilibrium fluctuations, $\nabla \delta\psi_{\rm neq}$, diverges in the inviscid limit.
Here, we show that the amplitude of the nonequilibrium fluctuation, $\delta\psi_{\rm neq}$, itself also diverges.
This divergence is demonstrated in Fig.~\ref{emfig2}(a), which plots the root-mean-square amplitude $\sqrt{\langle |\delta \psi_{\rm neq}|^2\rangle}$ as a function of $D_0$.
The data clearly show that the amplitude does not saturate at a constant value but instead increases monotonically as $D_0 \to 0$ without bounds.
The divergence, however, is remarkably slow, as is evident from the semi-logarithmic plot in Fig.~\ref{emfig2}(a).
Furthermore, as shown in Fig.~\ref{emfig2}(b), we find that the product $D_0 \langle |\delta \psi_{\rm neq}|^2\rangle$ vanishes in the inviscid limit.
This indicates that the divergence of $\langle |\delta \psi_{\rm neq}|^2\rangle$ is slower than $1/D_0$.
This scaling behavior is in contrast to that of the mean-squared gradient $\langle |\nabla \delta \psi_{\rm neq}|^2\rangle$, which diverges as $1/D_0$ to sustain the finite anomalous dissipation.

The fluctuation amplitude $\delta\psi_{\rm neq}$ is of significant physical interest.
It is directly proportional to the nonequilibrium Casimir pressure $p_{\rm neq} \propto \langle |\delta \psi_{\rm neq}|^2\rangle$~\cite{Kirkpatrick2013-ic, Kirkpatrick2015-rg, Kirkpatrick2016-ev}.
While this pressure has been a topic of extensive theoretical study~\cite{Wada2003-je, Kirkpatrick2013-ic, Kirkpatrick2015-rg, Kirkpatrick2016-ev, Croccolo2016-mn}, its experimental detection has remained a considerable challenge.
Our simulation results offer two perspectives on these experimental challenges.
First, the increase of $\sqrt{\langle |\delta \psi_{\rm neq}|^2\rangle}$ with decreasing $D_0$ is remarkably slow.
This suggests that simply choosing a fluid with a moderately smaller diffusivity is not an effective strategy to generate an easily observable pressure; rather, one would need to realize a physical system that corresponds to the extreme limit of vanishingly small $D_0$.
Second, we find that the widely used linear theory significantly overestimates $\sqrt{\langle |\delta \psi_{\rm neq}|^2\rangle}$ compared to our full nonlinear simulations, as shown in Fig.~\ref{emfig2}(a).
This finding underscores that for quantitatively accurate predictions, a theoretical framework incorporating nonlinear effects is essential.
A full theoretical understanding of this slow divergence remains a challenge for future work.

\newpage
%%%%%%%%%%%%%%%%%%%%%%%%%%%%%%%%%%%%%%

%\setcounter{figure}{0}
%\def\thefigure{S.\arabic{figure}}
%\setcounter{equation}{0}
%\def\theequation{A.\arabic{equation}}
\setcounter{equation}{0}
\setcounter{figure}{0}
\setcounter{table}{0}
\setcounter{page}{1}
% Use the number like (S1), (S2), etc.
\renewcommand{\thepage}{S\arabic{page}}  
\renewcommand{\thesection}{S\arabic{section}}   
\renewcommand{\thetable}{S\arabic{table}}   
\renewcommand{\thefigure}{S\arabic{figure}}
\renewcommand{\theequation}{S\arabic{equation}}
% \renewcommand{\bibnumfmt}[1]{[S#1]}
% \renewcommand{\citenumfont}[1]{S#1}
%%%%%%%%%%%%%%%%%%%%%%%%%%%%%%%%%%%%%%%

\begin{widetext}
\begin{center}
{\large \bf Supplemental Material for  \protect \\ 
  ``Dissipation anomaly in gradient-driven nonequilibrium steady states" }\\
\vspace*{0.3cm}
Hiroyoshi Nakano$^{1}$, and Yuki Minami$^{2}$
\\
\vspace*{0.1cm}
$^{1}${\small \it Institute for Solid State Physics, University of Tokyo, Kashiwa, Chiba 277-8581, Japan} \\
$^{2}${\small \it Faculty of Engineering, Gifu University, Yanagido, Gifu 501-1193, Japan} 
\end{center}

\section{I. Energetics of a fluctuating incompressible fluid}
This appendix outlines the derivation of the energy injection and dissipation rates for the scalar field $\psi$, following the framework of irreversible thermodynamics~\cite{degroot2013-ib}.
We consider an incompressible fluid, whose equations of motion are given by Eqs.~(2)--(3) in the main text.
For convenience, we restate the equation of motion:
\begin{align}
    \frac{\partial \bm{v}}{\partial t} = -\frac{1}{\rho_0} \nabla p + \nu_0 \nabla^2 \bm{v} - \nabla \cdot \bm{\Pi}_{\mathrm{ran}}, \\
    \frac{\partial \psi}{\partial t} + \nabla \cdot (\psi \bm{v}) = D_0 \nabla^2 \psi - \nabla \cdot \bm{J}_{\mathrm{ran}},
\end{align}
subjected to the incompressibility condition $\nabla \cdot \bm{v} = 0$.

\subsection{A. Expressions of energy injection and dissipation}
We begin by defining the central quantities of our energetic analysis: the injected power $\mathcal{P}_{\rm inj}$ and the dissipation rate $\mathcal{D}_{\rm diss}$.
The injected power is defined as the rate of energy exchange at the system's boundary, while the dissipation rate is defined as the rate of irreversible energy conversion within its volume.
Indeed, any change in the total free energy of the system, $F = \int d^d\bm{r} (\psi^2/2\chi)$, can be attributed to these two processes.
Its time evolution, $dF/dt$, is therefore described by the fundamental energy balance equation:
\begin{align}
    \frac{dF}{dt} = \mathcal{P}_{\rm inj} - \mathcal{D}_{\rm diss}.
\end{align}
In the following, we derive the mathematical expressions for $\mathcal{P}_{\rm inj}$ and $\mathcal{D}_{\rm diss}$ from the system's dynamics.
The dynamics of $\psi$ is rewritten as the continuity equation $\partial_t \psi + \nabla \cdot \bm{J}_{\psi} = 0$ with
\begin{align}
    \bm{J}_{\psi} := \psi \bm{v} - D_0 \nabla \psi + \bm{J}_{\rm ran}.
\end{align}
Using this flux, we calculate $dF/dt$ as:
\begin{align}
    \frac{dF}{dt} = - \int_V d^d \bm{r} \frac{\psi}{\chi} \nabla \cdot \bm{J}_{\psi}.
    \label{eq:appendix_tevol_F}
\end{align}
To separate this expression into boundary and bulk contributions, we decompose the total flux $\bm{J}_{\psi}$ into the reversible and irreversible components:
\begin{align}
    \bm{J}_{\psi} = \bm{J}_{\rm rev} + \bm{J}_{\rm irrev},
\end{align}
where $\bm{J}_{\rm rev} := \psi \bm{v}$ (an advective flux) and $\bm{J}_{\rm irrev} := - D_0 \nabla \psi + \bm{\mathcal{J}}_{\rm ran}$ (a diffusive and stochastic flux). 
Substituting this decomposition into Eq.~(\ref{eq:appendix_tevol_F}) and using integration by parts, we can analyze the contribution of each component.
The reversible term yields:
\begin{align}
& -\int_V d^d \bm{r} \left(\frac{\psi}{\chi}\right) \nabla \cdot \bm{J}_{\rm rev} = -\oint_{\partial V} d\bm{S} \cdot \left(\frac{\psi^2}{2\chi}\bm{v} \right),
\end{align}
where we have used the incompressibility condition $\nabla \cdot \bm{v} = 0$.
This confirms that the reversible advective process contributes only to energy exchange at the boundary.
In contrast, the irreversible term yields:
\begin{align}
& -\int_V d^d \bm{r} \left(\frac{\psi}{\chi}\right) \nabla \cdot \bm{J}_{\rm irrev} \nonumber \\
&= -\oint_{\partial V} d\bm{S} \cdot \left(\frac{\psi}{\chi} \bm{J}_{\rm irrev} \right) + \int_V d^d \bm{r} \bm{J}_{\rm irrev} \cdot \nabla \left(\frac{\psi}{\chi} \right).
\end{align}
This implies that irreversible processes contribute to both energy injection at the boundary and dissipation within the bulk.
By combining the results of these calculations, we can now write the expressions for the injected power and dissipation rate.
The injected power $\mathcal{P}_{\rm inj}$ is the sum of all terms that correspond to an energy flux across the boundary:
\begin{align}
    \mathcal{P}_{\rm inj} = - \oint_{\partial V} d \bm{S} \cdot \left(\frac{\psi^2}{2\chi}\bm{v} + \frac{\psi}{\chi} \bm{J}_{\rm irrev} \right).
    \label{eq:appendix_injected_power}
\end{align}
The dissipation rate $\mathcal{D}_{\rm diss}$ is identified as the remaining bulk term, which originates from the irreversible flux:
\begin{align}
    \mathcal{D}_{\rm diss} = - \int_V d^d \bm{r} \bm{J}_{\rm irrev} \cdot \nabla \left(\frac{\psi}{\chi} \right).
    \label{eq:appendix_dissiparion_rate}
\end{align}

\subsection{B. Application to the NESS with a uniform gradient}
In the main text, we decompose the solution of Eqs.~(2) and (3) into a mean profile and fluctuations: $\psi(\bm{r},t) = \langle \psi(y) \rangle + \delta \psi$, where $\langle \psi(y) \rangle = \psi_0 + G y$.
We consider a system where the fluctuations $\delta \psi$ and the velocity field $\bm{v}$ obey periodic boundary conditions in all directions.
First, let us derive the expression for the noise-averaged injected power $\langle \mathcal{P}_{\rm inj} \rangle$.
Substituting the decomposition of $\psi$ into Eq.(\ref{eq:appendix_injected_power}) and taking the average yields 
\begin{align}
    \langle \mathcal{P}_{\rm inj} \rangle &= - \left \langle \oint_{\partial V} d \bm{S} \cdot \left(\frac{\psi^2}{2\chi}\bm{v} + \frac{\psi}{\chi} \bm{J}_{\rm irrev} \right) \right \rangle, \nonumber \\
    &= - \oint_{\partial V} d \bm{S} \cdot \left(\frac{\langle \psi(y) \rangle^2}{2\chi} \langle \bm{v}\rangle + \frac{\langle \psi(y) \rangle}{\chi} \langle \delta \psi\bm{v}\rangle + \frac{1}{2\chi} \langle (\delta \psi)^2\bm{v} \rangle \right) - \oint_{\partial V} d \bm{S} \cdot \left(\frac{\langle \psi(y) \rangle}{\chi} \langle \bm{J}_{\rm irrev}\rangle + \frac{1}{\chi} \langle \delta \psi\bm{J}_{\rm irrev} \rangle \right).
\end{align}
The non-vanishing contributions arise because the mean profile $\langle \psi(y) \rangle = \psi_0 + G y$ is not periodic; it has a finite difference between the boundaries at $y=0$ and $y=L$.
Taking this non-periodicity into account, the work rate per unit volume, $\langle \dot{\varepsilon}_{\rm work}\rangle := \langle \mathcal{P}_{\rm inj} \rangle / V_d$, is calculated as:
\begin{align}
    \langle \dot{\varepsilon}_{\rm work}\rangle &= \frac{D_0}{\chi} G^2 - \frac{G}{\chi} \langle (\delta \psi) v^y\rangle \Big|_{\rm boundary}, \nonumber \\
    &=\frac{D_0}{\chi} G^2 - \frac{G}{\chi} \frac{1}{V_d} \int_V d^d \bm{r} \langle (\delta \psi) v^y\rangle. 
    \label{eq:appendix_work_rate_1}
\end{align}
In the second line, we have used the spatial homogeneity of the steady-state fluctuations to replace the boundary average with a volume average.
Furthermore, this expression can be simplified by considering the noise-averaged total flux, $\bm{J}_{\psi}$.
From its definition, we have:
\begin{align}
    \langle \bm{J}_{\psi}\rangle &= \Big\langle \big(\psi \bm{v} - D_0 \nabla \psi + \bm{J}_{\rm ran} \big)\Big \rangle, \nonumber \\
    &= \langle \psi \bm{v} \rangle - D_0 G.
\end{align}
Substituting this into Eq.~(\ref{eq:appendix_work_rate_1}) shows that the work rate $\langle \dot{\varepsilon}_{\rm work}\rangle$ can be written concisely in terms of the noise-averaged flux in the gradient direction:
\begin{align}
    \langle \dot{\varepsilon}_{\rm work}\rangle = - \frac{G}{\chi} \frac{1}{V_d} \int_V d^d\bm{r} \langle J^y_{\psi}\rangle.
\end{align}
Next, we study the noise-averaged dissipation rate $\langle \mathcal{D}_{\rm diss}\rangle$.
The calculation for this bulk quantity is more straightforward. By substituting the decomposition of $\psi$ and the definition of $\bm{J}_{\rm irrev}$ into Eq.~(\ref{eq:appendix_dissiparion_rate}), we obtain the dissipation rate per unit volume, $\langle \dot{\varepsilon}_{\rm diss}\rangle:= \langle \mathcal{D}_{\rm diss}\rangle / V_d$, as
\begin{align}
    \langle \dot{\varepsilon}_{\rm diss}\rangle &= - \frac{1}{V_d} \left\langle \int_V d^d \bm{r} \bm{J}_{\rm irrev} \cdot \nabla \left( \frac{\psi}{\chi} \right) \right\rangle, \nonumber \\
    &= \frac{D_0}{\chi} G^2 + \frac{1}{\chi} \frac{1}{V_d}\int_V d^d \bm{r} \biggl\langle \big(D_0 \nabla (\delta \psi) - \bm{J}_{\rm ran}\big) \cdot \nabla (\delta \psi)\biggr\rangle.
\end{align}

\subsection{C. Decomposition into equilibrium and nonequilibrium contributions}
To arrive at the final expressions used in the main text, we further decompose the fluctuation $\delta \psi$ into its equilibrium and non-equilibrium parts: $\delta \psi = \delta \psi_{\rm eq} + \delta \psi_{\rm neq}$. The equilibrium part $\delta \psi_{\rm eq}$ describes the thermal fluctuations that exist even at $G=0$ and obeys
\begin{align}
    \frac{\partial \psi_{\rm eq}}{\partial t} + \bm{v} \cdot \nabla \psi_{\rm eq} &= D_0 \nabla^2 \psi_{\rm eq} - \nabla \cdot \bm{J}_{\rm{ran}}.
    \label{eq:appendix_eom_psi_eq}
\end{align}
The non-equilibrium part $\delta\psi_{\rm neq}$ is the additional fluctuation induced by the gradient, which follows
\begin{align}
    \frac{\partial \psi_{\rm neq}}{\partial t} + v_y G + \bm{v} \cdot \nabla \psi_{\rm neq} &= D_0 \nabla^2 \psi_{\rm neq}.
    \label{eq:appendix_eom_psi_neq}
\end{align}
This decomposition allows us to isolate the energetic contributions that arise purely from the non-equilibrium driving.
For the work rate, the correlation $\langle (\delta \psi) v^y\rangle$ becomes $\langle (\delta \psi_{\rm eq}) v^y\rangle + \langle (\delta \psi_{\rm neq}) v^y\rangle$.
Here, $\langle (\delta \psi_{\rm eq}) v^y\rangle$ must vanish due to time-reversal symmetry in equilibrium, i.e., $\langle (\delta \psi_{\rm eq}) v^y\rangle = 0$.
Thus, the work rate depends solely on the non-equilibrium fluctuations:
\begin{align}
    \langle \dot{\varepsilon}_{\rm work}\rangle =\frac{D_0}{\chi} G^2 - \frac{G}{\chi} \frac{1}{V_d} \int_V d^d \bm{r} \langle (\delta \psi_{\rm neq}) v^y\rangle. 
    \label{eq:appendix_wr_final}
\end{align}
For the dissipation rate, we use the fact that in the equilibrium state ($G=0$), the system is stationary, meaning the total average dissipation must be zero.
This leads to the condition on $\delta \psi_{\rm eq}$:
\begin{align}
    D_0 \int_V d^d \bm{r} \left\langle |\nabla (\delta \psi_{\rm eq})|^2 \right\rangle = \int_V d^d \bm{r} \left\langle \bm{J}_{\rm ran} \cdot \nabla (\delta \psi_{\rm eq})\right\rangle.
\end{align}
Consequently, when we evaluate the total dissipation rate for the NESS, all terms involving only $\delta \psi_{\rm eq}$ cancel out.
The remaining part is given by:
\begin{align}
    \langle \dot{\varepsilon}_{\rm diss}\rangle = \frac{D_0}{\chi} G^2 + \frac{D_0}{\chi} \frac{1}{V_d} \int_V d^d \bm{r} \langle |\nabla(\delta\psi_{\rm neq})|^2 \rangle.
    \label{eq:appendix_dr_final}
\end{align}
This final expression corresponds to Eq.~(7) in the main text.

\section{II. Self-consistent mode-coupling theory}
This appendix outlines the details of the theoretical calculation of the energy dissipation rate $\langle \dot{\varepsilon}_{\rm diss}\rangle$ based on the self-consistent mode-coupling theory (MCT).
In the steady state, this rate is equal to the work rate $\langle \dot{\varepsilon}_{\rm work}\rangle$.
Since the expression for the work rate [Eq.~(\ref{eq:appendix_wr_final})] allows for a more straightforward calculation than the one for the dissipation rate [Eq.~(\ref{eq:appendix_dr_final})], we will focus on evaluating $\langle \dot{\varepsilon}_{\rm work}\rangle$ in this appendix.
For convenience, we restate here the exact expression for the work rate:
\begin{align}
    \langle \dot{\varepsilon}_{\rm work}\rangle = \frac{D_0}{\chi} G^2 - \frac{G}{\chi} \langle (\delta \psi_{\rm neq}) v^y\rangle.
\label{eq:obj_appendix_b}
\end{align}

\subsection{A. Governing equation and formal solution}
To calculate the work rate, we must evaluate the cross-correlation $\langle (\delta \psi_{\rm neq}) v^y\rangle$.
Our analysis therefore begins with the equation of motion for the non-equilibrium fluctuations of the scalar field, $\delta\psi_{\rm neq}$, Eq.~(\ref{eq:appendix_eom_psi_neq}).
In Fourier space, this equation is given by:
\begin{align}
    (-i\omega + D_0 k^2) \delta\psi_{\rm neq}(\bm{k},\omega) = -G v^y(\bm{k},\omega) - \mathcal{V}_{\rm NL}[\bm{v}, \delta\psi_{\rm neq}],
    \label{eq:appendix_eom_psi}
\end{align}
where $\mathcal{V}_{\rm NL}[\bm{v}, \delta\psi_{\rm neq}]$ is the nonlinear advection term:
\begin{align}
    \mathcal{V}_{\rm NL}[\bm{v}, \delta\psi_{\rm neq}] = i \int \frac{d^d\bm{q}d\Omega}{(2\pi)^{d+1}} \bm{q} \cdot \bm{v}(\bm{k}-\bm{q}, \omega-\Omega) \delta\psi_{\rm neq}(\bm{q},\Omega).
\end{align}
We introduce the bare propagator, $\mathcal{G}_0(\bm{k},\omega)$, which describes the dynamics of the system in the absence of the nonlinear term.
It is defined as the inverse of the linear operator on the left-hand side of Eq.~(\ref{eq:appendix_eom_psi}):
\begin{align}
    \mathcal{G}_0(\bm{k},\omega) := \frac{1}{- i \omega + D_0 k^2}.
\end{align}
Using the bare propagator, we can write a formal solution to Eq.~(\ref{eq:appendix_eom_psi}) as an integral equation:
\begin{align}
    \delta\psi_{\rm neq}(\bm{k},\omega) = \mathcal{G}_0(\bm{k},\omega) \Big( -G v^y(\bm{k},\omega) - \mathcal{V}_{\rm NL}[\bm{v}, \delta\psi_{\rm neq}] \Big).
    \label{eq:appendix_fs_psi}
\end{align}
While Eq.~(\ref{eq:appendix_fs_psi}) is not a closed-form solution, it serves as the formal starting point for the perturbative analysis in the subsequent sections.

\subsection{B. Linear approximation}
As a first step, we calculate $\langle \dot{\varepsilon}_{\rm work}\rangle$ within the linear approximation.
Neglecting the nonlinear term $\mathcal{V}_{\rm NL}$ simplifies the formal solution Eq.~(\ref{eq:appendix_fs_psi}) and yields the following explicit solution:
\begin{align}
    \delta\psi^{(0)}_{\rm neq} = \mathcal{G}_0(\bm{k},\omega) (-G v^y(\bm{k},\omega)).
\end{align}
Substituting this linear solution into Eq.~(\ref{eq:obj_appendix_b}), we obtain $\langle \dot{\varepsilon}_{\rm work}\rangle$ in the linear approximation, denoted as $\langle \dot{\varepsilon}_{\rm work}\rangle_{\rm lin}$:
\begin{align}
    \langle \dot{\varepsilon}_{\rm work}\rangle_{\rm lin} &= \frac{D_0}{\chi} G^2 - \frac{G}{\chi} \langle \delta\psi^{(0)}_{\rm neq} v^y \rangle, \nonumber \\
    &= \frac{D_0}{\chi} G^2 - \frac{G}{\chi} \int \frac{d^d\bm{k}d\omega}{(2\pi)^{d+1}} \langle \delta\psi^{(0)}_{\rm neq}(\bm{k},\omega) v^y(-\bm{k},-\omega) \rangle, \nonumber \\
    &= \frac{D_0}{\chi} G^2 - \frac{G}{\chi} \int \frac{d^d\bm{k}d\omega}{(2\pi)^{d+1}} \langle \mathcal{G}_0(\bm{k},\omega) (-G v^y(\bm{k},\omega)) v^y(-\bm{k},-\omega) \rangle, \nonumber \\
    &= \frac{D_0}{\chi} G^2 + \frac{G^2}{\chi} \int \frac{d^d\bm{k}d\omega}{(2\pi)^{d+1}} \mathcal{G}_0(\bm{k},\omega) C_{yy}(\bm{k},\omega),
    \label{eq:appendix_wr_up_to_linear}
\end{align}
where $C_{yy}(\bm{k},\omega)$ is the velocity correlation function, which is explicitly given by (see Sec.~III)
\begin{align}
    C_{ab}(\bm{k}, \omega) = 2\frac{k_B T}{\rho_0} \nu_0 \frac{\delta_{ab} \bm{k}^2 - k_a k_b}{\omega^2 + \nu_0^2 \bm{k}^4}.
\end{align}
The frequency integration can be performed analytically, and we obtain the expression for $\langle \dot{\varepsilon}_{\rm work}\rangle_{\rm lin}$:
\begin{align}
    \langle \dot{\varepsilon}_{\rm work}\rangle_{\rm lin} = \frac{D_0}{\chi} G^2 + \frac{k_B T}{\rho_0} \frac{G^2}{\chi(\nu_0 + D_0)} \int \frac{d^d\bm{k}}{(2\pi)^d} \frac{k^2-k_y^2}{k^4}.
    \label{eq:appendix_wr_lin}
\end{align}
This is Eq.~(12) in the main text.

\subsection{C. Perturbation theory: one-loop correction}
To improve upon the linear approximation, we incorporate the effects of the nonlinear term $\mathcal{V}_{\rm NL}$ using perturbation theory~\cite{Mazenko2006}.
We return to the formal solution [Eq.~(\ref{eq:appendix_fs_psi})] and generate a perturbation series for $\delta\psi_{\rm neq}$ by iterative substitution:
\begin{align}
    \delta\psi_{\rm neq} = \underbrace{\mathcal{G}_0 (-G v^y)}_{\delta\psi^{(0)}_{\rm neq}} \underbrace{- \mathcal{G}_0 \mathcal{V}_{\rm NL}[\bm{v}, \delta\psi^{(0)}_{\rm neq}]}_{\delta\psi^{(1)}_{\rm neq}} + \underbrace{\mathcal{G}_0 \mathcal{V}_{\rm NL}[\bm{v}, \delta\psi^{(1)}_{\rm neq}]}_{\delta\psi^{(2)}_{\rm neq}} + \dots
\end{align}
Substituting this series into Eq.~(\ref{eq:obj_appendix_b}), we obtain a series expansion for $\langle \dot{\varepsilon}_{\rm work}\rangle$.
\begin{align}
    \langle \dot{\varepsilon}_{\rm work}\rangle = \underbrace{\frac{D_0}{\chi} G^2 - \frac{G}{\chi} \langle \delta\psi^{(0)}_{\rm neq} v^y \rangle}_{\rm linear} \underbrace{- \frac{G}{\chi} \langle \delta\psi^{(2)}_{\rm neq} v^y \rangle}_{\rm 1-loop} - \cdots,
    \label{eq:appendix_work_expanded}
\end{align}
where terms with an odd number of velocity fields vanish due to the Gaussian statistics of the velocity field ($\langle \bm{v} \rangle = \bm{0}$).
The first two terms correspond to the contribution from the linear theory [Eq.~(\ref{eq:appendix_wr_lin})].
The last term is the leading-order nonlinear contribution, also known as the one-loop correction.
A direct calculation shows that this term can be expressed as:
\begin{align}
    - \frac{G}{\chi} \langle \delta\psi^{(2)}_{\rm neq} v^y \rangle = \frac{G^2}{\chi} \int \frac{d^d\bm{k}d\omega}{(2\pi)^{d+1}} \mathcal{G}_0(\bm{k},\omega) \Sigma^{(2)}(\bm{k},\omega) \mathcal{G}_0(\bm{k},\omega) C_{yy}(\bm{k},\omega),
    \label{eq:appendix_wr_one_loop}
\end{align}
where $\Sigma^{(2)}(\bm{k},\omega)$ is the one-loop self-energy:
\begin{align}
    \Sigma^{(2)}(\bm{k}, \omega) = - \int \frac{d^d\bm{q} d\Omega}{(2\pi)^{d+1}} \sum_{a,b} q_a k_b C_{ab}(\bm{k}-\bm{q}, \omega-\Omega) \mathcal{G}_0(\bm{q}, \Omega).
    \label{eq:appendix_self_energy_oneloop}
\end{align}
By substituting the expressions for the linear [Eq.~(\ref{eq:appendix_wr_up_to_linear})] and one-loop [Eq.~(\ref{eq:appendix_wr_one_loop})] contributions back into the expansion [Eq.~(\ref{eq:appendix_work_expanded})], we obtain $\langle \dot{\varepsilon}_{\rm work}\rangle$ up to the one-loop order:
\begin{align}
    \langle \dot{\varepsilon}_{\rm work}\rangle \approx \frac{D_0}{\chi} G^2 + \frac{G^2}{\chi} \int \frac{d^d\bm{k}d\omega}{(2\pi)^{d+1}} \left(\mathcal{G}_0(\bm{k},\omega) + \mathcal{G}_0(\bm{k},\omega) \Sigma^{(2)}(\bm{k},\omega) \mathcal{G}_0(\bm{k},\omega) \right)C_{yy}(\bm{k},\omega).
    \label{eq:appendix_wr_up_to_oneloop}
\end{align}
In principle, this perturbation series can be extended to arbitrarily high orders.

\subsection{D. Self-consistent mode-coupling theory (MCT)}
A simple truncation of the perturbation series at the one-loop level may not be sufficiently accurate.
To obtain a more powerful result that goes beyond the simple perturbation theory, we introduce the self-consistent MCT. 
The central quantity in this framework is the full propagator $\mathcal{G}(\bm{k},\omega)$.
It is defined as the effective response function that encapsulates the system's total response to the driving force in the following linear form:
\begin{align}
    \delta\psi_{\rm neq}(\bm{k},\omega) = \mathcal{G}(\bm{k},\omega) \Big( -G v^y(\bm{k},\omega)\Big).
    \label{eq:appendix_fs_with_dressed_propagator}
\end{align}
This definition allows the work rate [Eq.~(\ref{eq:obj_appendix_b})] to be written compactly in terms of $\mathcal{G}$:
\begin{align}
    \langle \dot{\varepsilon}_{\rm work}\rangle = \frac{D_0}{\chi} G^2 + \frac{G^2}{\chi}\int \frac{d^d\bm{k}d\omega}{(2\pi)^{d+1}} \mathcal{G}(\bm{k},\omega) C_{yy}(\bm{k},\omega).
    \label{eq:appendix_work_in_G}
\end{align}
By demanding that this effective description must be consistent with the perturbation theory, we can determine the structure of $\mathcal{G}$.
Indeed, comparing Eq.~(\ref{eq:appendix_work_in_G}) with the one-loop result [Eq.~(\ref{eq:appendix_wr_up_to_oneloop})] leads to the one-loop approximation for $\mathcal{G}$ as:
\begin{align}
    \mathcal{G}(\bm{k}, \omega) \approx \mathcal{G}_0(\bm{k}, \omega) + \mathcal{G}_0(\bm{k}, \omega) \Sigma^{(2)}(\bm{k}, \omega) \mathcal{G}_0(\bm{k}, \omega).
    \label{eq:appendix_one-loop_G}
\end{align}
Furthermore, continuing this expansion to higher orders, $\mathcal{G}$ is generally expressed as the following infinite series constructed from the full self-energy $\Sigma$:
\begin{align}
    \mathcal{G} = \mathcal{G}_0 + \mathcal{G}_0 \Sigma \mathcal{G}_0 + \mathcal{G}_0 \Sigma \mathcal{G}_0 \Sigma \mathcal{G}_0 + \dots
\end{align}
Formally summing this infinite series yields the well-known Dyson equation:
\begin{align}
    \mathcal{G}(\bm{k},\omega) = \frac{1}{\mathcal{G}_0^{-1}(\bm{k},\omega) - \Sigma(\bm{k},\omega)} = \frac{1}{- i \omega + D_0 k^2 - \Sigma(\bm{k},\omega)}.
    \label{eq:appendix_dyson_eq}
\end{align}

We here introduce the one-loop resummation approximation.
It is constructed by replacing $\Sigma$ in the Dyson equation [Eq.~(\ref{eq:appendix_dyson_eq})] with the one-loop self-energy $\Sigma^{(2)}$ [Eq.~(\ref{eq:appendix_self_energy_oneloop})].
This procedure yields 
\begin{align}
    \mathcal{G}(\bm{k},\omega) \approx \frac{1}{- i \omega + D_0 k^2 - \Sigma^{(2)}(\bm{k},\omega)}.
    \label{eq:one-loop_resummation}
\end{align}
This approximation is more powerful than the simple one-loop perturbation approximation [Eq.~(\ref{eq:appendix_one-loop_G})] because it resums an infinite series of diagrams where the one-loop self-energy block is repeatedly inserted.

The self-consistent MCT further improves upon the one-loop resummation.
In this framework, the self-energy $\Sigma$ itself is determined self-consistently by replacing the bare propagator $\mathcal{G}_0$ in the one-loop self-energy $\Sigma^{(2)}$ [Eq.~(\ref{eq:appendix_self_energy_oneloop})] with the dressed propagator $\mathcal{G}$:
\begin{align}
    \Sigma(\bm{k}, \omega) \approx - \int \frac{d^d\bm{q} d\Omega}{(2\pi)^{d+1}} \sum_{a,b} q_a k_b C_{ab}(\bm{k}-\bm{q}, \omega-\Omega) \mathcal{G}(\bm{q}, \Omega).
    \label{eq:appendix_sc_self_energy}
\end{align}
By substituting this self-energy into the Dyson equation [Eq.~(\ref{eq:appendix_dyson_eq})], a much larger class of diagrams is resummed compared to the one-loop resummation, leading to a more accurate result.
However, the expression for $\Sigma$ [Eq.~(\ref{eq:appendix_sc_self_energy})] now depends on the unknown propagator $\mathcal{G}$ itself.
Therefore, it must be solved simultaneously (or self-consistently) with the Dyson equation [Eq.~(\ref{eq:appendix_dyson_eq})].

\subsection{E. The approximate solution to the self-consistent MCT}
We now seek an approximate analytical solution to the self-consistent MCT equations [Eqs.~(\ref{eq:appendix_dyson_eq}) and (\ref{eq:appendix_sc_self_energy})] in the long-wavelength ($\bm{k} \to \bm{0}$) and low-frequency ($\omega \to 0$) limit.

First, we rewrite $\mathcal{G}$ by using the static approximation for the self-energy.
Assuming that the frequency dependence of $\Sigma(\bm{k}, \omega)$ is negligible for the steady-state properties of interest, we replace it with its static value, $\Sigma(\bm{k}, \omega=0)$:
\begin{align}
    \mathcal{G}(\bm{k},\omega) = \frac{1}{- i \omega + D_0 k^2 - \Sigma(\bm{k}, \omega)} \approx \frac{1}{- i \omega + D_0 k^2 - \Sigma(\bm{k}, \omega=0)}.
\end{align}
Here, it is useful to define a renormalized diffusion coefficient $D_R(\bm{k})$:
\begin{align}
    D_R(\bm{k}) k^2 := D_0 k^2 - \Sigma(\bm{k}, \omega=0).
    \label{eq:appendix_def_DR}
\end{align}
Under this definition, $\mathcal{G}$ is written compactly as
\begin{align}
    \mathcal{G}(\bm{k},\omega) \approx \frac{1}{- i \omega + D_R(\bm{k}) k^2}.
    \label{eq:appendix_G_static_approximation}
\end{align}
Given this approximation for $\mathcal{G}$, we can proceed with the calculation of the self-energy [Eq.~(\ref{eq:appendix_sc_self_energy})].
After performing the frequency integration, the static self-energy is given by
\begin{align}
    \Sigma(\bm{k}, \omega=0) = - \frac{k_B T}{\rho_0} \int \frac{d^d\bm{q}}{(2\pi)^d} \frac{\bm{q}^2 \bm{k}^2 - (\bm{q} \cdot \bm{k})^2}{\bm{p}^2 (\nu_0 \bm{p}^2 + D_R(\bm{q})\bm{q}^2)},
    \label{eq:appendix_static_self_energy}
\end{align}
where $\bm{p}:=\bm{k} - \bm{q}$.
From the definition of $D_R(\bm{k})$ [Eq.~(\ref{eq:appendix_def_DR})], we immediately find that this is the self-consistent integral equation for $D_R(\bm{k})$:
\begin{align}
    (D_0 - D_R(\bm{k})) k^2 = - \frac{k_B T}{\rho_0} \int \frac{d^d\bm{q}}{(2\pi)^d} \frac{\bm{q}^2 \bm{k}^2 - (\bm{q} \cdot \bm{k})^2}{\bm{p}^2 (\nu_0 \bm{p}^2 + D_R(\bm{q})\bm{q}^2)}.
    \label{eq:appendix_sc_eq}
\end{align}
The wavenumber integral in Eq.~(\ref{eq:appendix_sc_eq}) cannot be performed exactly.
Therefore, we approximately evaluate it in the long-wavelength limit ($\bm{k} \to \bm{0}$).
By performing a careful integral evaluation, detailed in Sec.~IIG, we derive the following analytical expression:
\begin{align}
    \int \frac{d^d\bm{q}}{(2\pi)^d} \frac{\bm{q}^2 \bm{k}^2 - (\bm{q} \cdot \bm{k})^2}{\bm{p}^2 (\nu_0 \bm{p}^2 + D_R(\bm{q})\bm{q}^2)} \approx \frac{S_{d-1}}{(2\pi)^d} \frac{d-1}{d} \frac{k^2}{\nu_0 + D_R(\bm{k})} \int_k^\Lambda q^{d-3} dq \qquad {\rm for} \quad \bm{k} \to \bm{0}, 
    \label{eq:appendix_sigma_approx}
\end{align}
where $S_{d-1}$ is the surface area of a $(d-1)$-dimensional unit sphere and $\Lambda = 2\pi/a_{uv}$ is the ultraviolet cutoff.
Substituting this result into the self-consistent equation for $D_R(\bm{k})$ [Eq.~(\ref{eq:appendix_sc_eq})], the complex integral equation reduces to a simple algebraic equation for $D_R(\bm{k})$:
\begin{align}
    (D_0 - D_R(\bm{k})) k^2 = - \frac{k_B T}{\rho_0} \frac{S_{d-1}}{(2\pi)^d} \frac{d-1}{d} \frac{k^2}{\nu_0 + D_R(\bm{k})} \int_k^\Lambda q^{d-3} dq.
\end{align}
Solving this quadratic equation for $D_R(\bm{k})$ and taking the physically meaningful (positive) root, we obtain the expression:
\begin{align}
    D_R(\bm{k}) = \frac{D_0 - \nu_0}{2} + \frac{D_0 + \nu_0}{2} \sqrt{1 + 4 \frac{k_B T}{\rho_0} \frac{S_{d-1}}{(2\pi)^d} \frac{d-1}{d} \frac{1}{(\nu_0 + D_0)^2} \int_{k}^{\Lambda} q^{d-3} dq } \;.
    \label{eq:appendix_DR_final}
\end{align}
This expression corresponds to Eq.~(14) in the main text.

\subsection{F. The work rate in the self-consistent MCT}
We now calculate the final expression for the work rate using the self-consistent propagator $\mathcal{G}$ derived in the previous sections.
Let us recall the formal expression for the work rate in terms of $\mathcal{G}$ [Eq.~(\ref{eq:appendix_work_in_G})]:
\begin{align}
    \langle \dot{\varepsilon}_{\rm work}\rangle = \frac{D_0}{\chi} G^2 + \frac{G^2}{\chi}\int \frac{d^d\bm{k}d\omega}{(2\pi)^{d+1}} \mathcal{G}(\bm{k},\omega) C_{yy}(\bm{k},\omega).
\end{align}
To evaluate this integral, we use the approximate form of $\mathcal{G}$, which is given by [Eq.~(\ref{eq:appendix_G_static_approximation})]
\begin{align}
    \mathcal{G}(\bm{k},\omega) \approx \frac{1}{-i\omega + D_R(\bm{k})k^2},
\end{align}
where the renormalized diffusion coefficient $D_R(\bm{k})$ is calculated as Eq.~(\ref{eq:appendix_DR_final}).
Substituting this into the work rate expression and performing the frequency integration yields:
\begin{align}
    \langle \dot{\varepsilon}_{\rm work}\rangle_{\rm sc} = \frac{D_0}{\chi} G^2 + \frac{G^2}{\chi} \frac{k_B T}{\rho_0} \int \frac{d^d\bm{k}}{(2\pi)^d} \frac{1}{\nu_0 + D_R(\bm{k})} \frac{k^2-k_y^2}{k^4}.
    \label{eq:appendix_work_final_sc}
\end{align}
This is the final expression for the work rate within our MCT framework, corresponding to Eq.~(13) in the main text.

\subsection{G. Evaluation of the wavenumber integral in Eq.~(\ref{eq:appendix_sc_eq})}
\label{sec:Evaluation of the wavenumber integral}
In this subsection, we present the detailed evaluation of the wavenumber integral appearing in the self-consistent equation for $D_R(\bm{k})$ [Eq.~(\ref{eq:appendix_sc_eq})].
For convenience, we restate the integral, which we denote as $I(\bm{k})$:
\begin{align}
    I(\bm{k}) := \int \frac{d^d\bm{q}}{(2\pi)^d} \frac{q^2k^2 - (\bm{q}\cdot\bm{k})^2}{\bm{p}^2(\nu_0\bm{p}^2 + D_R(\bm{q})q^2)},
\end{align}
where $\bm{p} = \bm{k}-\bm{q}$.
To make the integral tractable, we first assume that the $\bm{q}$-dependence of $D_R(\bm{q})$ inside the integral is weak, allowing us to replace it with its limiting value $D_R(\bm{q}=\bm{0})$.
With this simplification, the integral is given by:
\begin{align}
I(\bm{k}) &= \int \frac{d^d\bm{q}}{(2\pi)^d} \frac{q^2k^2 - (\bm{q}\cdot\bm{k})^2}{\bm{p}^2(\nu_0\bm{p}^2 + D_R(\bm{0})q^2)}.
\end{align}
To proceed, we decompose the integration domain into two regions, $k>q$ and $k<q$: 
\begin{align}
I(\bm{k}) = I_{k>q}(\bm{k}) + I_{k<q}(\bm{k}).
\end{align}

\paragraph*{Contribution from the region $k>q$:}
First, we consider the contribution from the region $k>q$
In this region, we approximate $\bm{p} = \bm{k}-\bm{q} \approx \bm{k}$.
The integral becomes:
\begin{align}
    I_{k>q}(\bm{k}) \approx \int_{k>q} \frac{d^d\bm{q}}{(2\pi)^d} \frac{q^2k^2 - (\bm{q}\cdot\bm{k})^2}{k^2(\nu_0 k^2 + D_R(\bm{0})q^2)}.
\end{align}
After performing the angular integration and approximating the denominator for small $\bm{q}$, we find:
\begin{align}
I_{k>q}(\bm{k}) &\approx \frac{S_{d-1}}{(2\pi)^d} \frac{d-1}{d} \int_0^k dq \frac{q^{d+1}}{\nu_0 k^2 + D_R(\bm{0}) q^2} \approx \frac{S_{d-1}}{(2\pi)^d} \frac{d-1}{d\nu_0(d+2)} k^d,
\end{align}
where $S_{d-1}$ represents the surface area of a $(d-1)$-dimensional unit sphere.
From this expression, we find that $I_{k>q}(\bm{k})$ scales as $k^d$.

\paragraph*{Contribution from the region $k<q$:}
Next, we consider the contribution from the region $k<q$.
In this region, we approximate $\bm{p} = \bm{k} - \bm{q} \approx - \bm{q}$.
This approximation is applied to the denominator, while the numerator is treated exactly to preserve its angular structure:
\begin{align}
    I_{k<q}(\bm{k}) &\approx \int_{k<q} \frac{d^d\bm{q}}{(2\pi)^d} \frac{q^2k^2 - (\bm{q}\cdot\bm{k})^2}{q^2(\nu_0 q^2 + D_R(\bm{0})q^2)} = \frac{1}{\nu_0 + D_R(\bm{0})} \int_{k<q} \frac{d^d\bm{q}}{(2\pi)^d} \frac{q^2k^2 - (\bm{q}\cdot\bm{k})^2}{q^4}.
\end{align}
Performing the angular integration as before yields:
\begin{align}
    I_{k<q}(\bm{k}) = \frac{S_{d-1}}{(2\pi)^d} \frac{d-1}{d} \frac{1}{\nu_0 + D_R(\bm{0})} k^2 \int_k^\Lambda q^{d-3} dq.
\end{align}
The $\bm{k}$-dependence of $I_{k<q}(\bm{k})$ depends on the dimension $d$.
For $d=2$, $I_{k<q}(\bm{k})$ is explicitly written as
\begin{align}
    I_{k<q}(\bm{k}) = \frac{1}{4\pi} \frac{1}{\nu_0 + D_R(\bm{0})} k^2 \Big(\log \Lambda - \log k\Big),
\end{align}
which implies that the integral scales as $k^2 \log k$ as $\bm{k} \to \bm{0}$.
For $d=3$, $I_{k<q}(\bm{k})$ is explicitly written as
\begin{align}
    I_{k<q}(\bm{k}) = \frac{1}{3\pi^2} \frac{1}{\nu_0 + D_R(\bm{0})} k^2 \Big(\Lambda - k\Big),
\end{align}
which implies that the integral scales as $k^2$ as $\bm{k} \to \bm{0}$.

\paragraph*{Dominant contribution and final result:}
We summarize the results of the above analysis for the contributions from the regions $k>q$ and $k<q$.
%the IR and UV contributions.
The $k>q$ contribution scales as:
\begin{align}
     I_{k>q}(\bm{k}) \sim k^d \qquad {\rm for \ any \ }d.
\end{align}
In contrast, the $k<q$ contribution scales as:
\begin{align}
    I_{k<q}(\bm{k}) \sim
        \begin{cases}
            k^2 \log k & {\rm for \ } d = 2, \\
            k^2 & {\rm for \ } d = 3.
        \end{cases}
\end{align}
From this comparison, it is clear that $I_{k<q}(\bm{k})$ provides the dominant contribution to the integral in the long-wavelength limit for $d \geq 2$.
We therefore approximate the full integral $I(\bm{k})$ with this dominant part.
Finally, we replace the constant $D_R(\bm{0})$, which was used for the convenience of integration, with $D_R(\bm{k})$.
This yields our final approximation for the integral:
\begin{align}
    I(\bm{k}) \approx I_{k<q}(\bm{k}) = \frac{S_{d-1}}{(2\pi)^d} \frac{d-1}{d} \frac{k^2}{\nu_0 + D_R(\bm{k})} \int_k^\Lambda q^{d-3} dq.
    \label{eq:appendix_integral_final_approx}
\end{align}

\section{III. Justification of the incompressible limit}
In the main text, we analyzed the system's behavior using an incompressible fluid model.
However, our numerical simulations were performed using a compressible fluid model in the regime of a large, finite speed of sound, $c_T$.
This appendix serves to justify this approach.
We explicitly demonstrate that in this regime, the contributions from longitudinal modes to the energy dissipation rate become negligible.
This confirms that the compressible model with large $c_T$ accurately captures the behaviors of the incompressible model.

\subsection{A. Compressible fluctuating hydrodynamics}
In the main text, we considered a scalar field $\psi$ advected by a purely incompressible velocity field ($\nabla \cdot \bm{v} = 0$).
In this appendix, we consider the compressible version of this model.
The system is then described by the total mass density $\rho(\bm{r},t)$, the fluid velocity $\bm{v}(\bm{r},t)$, and the scalar field $\psi(\bm{r},t)$.
This scalar field can be interpreted, for example, as the mass fraction (concentration) of one component in a binary fluid.
Its dynamics are governed by the fluctuating hydrodynamic equations~\cite{Donev2014-jy, Donev2015-tm}:
\begin{align}
    & \frac{\partial \rho}{\partial t} + \nabla \cdot (\rho \bm{v}) = 0, \label{eq:app_cont} \\[2pt]
    & \rho \frac{\partial \bm{v}}{\partial t} = - \nabla p + \eta_0 \nabla^2 \bm{v} + [\zeta_0 + (1-\tfrac{2}{d})\eta_0]\nabla(\nabla \cdot \bm{v}) - \nabla \cdot \bm{\Pi}_{\rm{ran}}, \label{eq:app_ns} \\[2pt]
    & \rho \left(\frac{\partial \psi}{\partial t} + \bm{v} \cdot \nabla \psi \right) = \nabla \cdot (\rho D_0 \nabla \psi) - \nabla \cdot (\rho \bm{J}_{\mathrm{ran}}). \label{eq:app_cd}
\end{align}
Here, $\eta_0$ is the shear viscosity, $\zeta_0$ is the bulk viscosity, and $D_0$ is the diffusion coefficient.
We assume a simple barotropic equation of state, $p = c_T^2 \rho$, where $c_T$ is the isothermal sound speed.
The stochastic fluxes $\bm{\Pi}_{\mathrm{ran}}$ and $\bm{J}_{\mathrm{ran}}$ represent thermal noise, obeying the fluctuation–dissipation theorem:
\begin{align}
    & \left \langle \Pi_{\mathrm{ran}}^{ab}(\bm{r},t) \Pi_{\mathrm{ran}}^{cd}(\bm{r}',t') \right \rangle = 2 k_B T \delta(\bm{r}-\bm{r}')\delta(t-t') \Big[\eta_0(\delta^{ac}\delta^{bd}+\delta^{ad}\delta^{bc}) + (\zeta_0 - \tfrac{2}{d}\eta_0)\delta^{ab}\delta^{cd}\Big], \\
    & \left\langle J_{\mathrm{ran}}^{a}(\bm{r},t) J_{\mathrm{ran}}^{b}(\bm{r}',t') \right\rangle
    = \frac{2 k_B T \chi}{\rho} D_0\, \delta^{ab}\delta(\bm{r}-\bm{r}')\delta(t-t').
\end{align}
These equations provide a complete description of the compressible system.

\subsection{B. Dynamic and Static Correlations}
By linearizing Eqs.~(\ref{eq:app_cont}) and (\ref{eq:app_ns}) around the equilibrium state ($\rho = \rho_0, \bm{v} = 0$), we can calculate the dynamic correlation functions of the hydrodynamic modes.
These correlation functions are defined in Fourier space via:
\begin{align}
    \Big\langle \delta\rho(\bm{k},\omega) \delta\rho(\bm{k}',\omega') \Big\rangle &= (2\pi)^{d+1} \delta(\bm{k} + \bm{k}') \delta(\omega + \omega') C_{\rho\rho}(\bm{k},\omega), \\
    \Big\langle v_a(\bm{k},\omega) v_b(\bm{k}',\omega') \Big\rangle &= (2\pi)^{d+1} \delta(\bm{k} + \bm{k}') \delta(\omega + \omega') C_{ab}(\bm{k},\omega).
\end{align}
Here, $\delta\rho = \rho - \rho_0$ is the density fluctuation.
A straightforward calculation yields the explicit forms of these spectra. The velocity correlation $C_{ab}$ is decomposed into longitudinal (L) and transverse (T) components:
\begin{align}
    C_{ab}(\bm{k},\omega) &= C^{L}(k, \omega) P^L_{ab}(\hat{\bm{k}}) + C^{T}(k, \omega) P^T_{ab}(\hat{\bm{k}}),
\end{align}
where $P^L_{ab}(\hat{\bm{k}}) = k_a k_b / k^2$ and $P^T_{ab}(\hat{\bm{k}}) = \delta_{ab} - k_a k_b / k^2$ are the projection operators onto each component.
The resulting spectra are~\cite{chaikin1995-pr}:
\begin{align}
    % Longitudinal Velocity
    C^{L}(k, \omega) &= \frac{2 k_B T}{\rho_0} \frac{\Gamma_0 k^2 \omega^2}{\Gamma_0^2 k^4 \omega^2 + (c_T^2 k^2 - \omega^2)^2}, \\[5pt]
    % Transverse Velocity
    C^{T}(k, \omega) &= \frac{2 k_B T}{\rho_0} \frac{\nu_0 k^2}{\omega^2 + (\nu_0 k^2)^2}, \\[5pt]
    % Density
    C_{\rho\rho}(\bm{k},\omega) &= 2 k_B T \rho_0 \frac{\Gamma_0 k^4}{\Gamma_0^2 k^4 \omega^2 + (c_T^2 k^2 - \omega^2)^2}.
\end{align}
Here, $\nu_0 = \eta_0 / \rho_0$ is the kinematic viscosity, and $\Gamma_0 = [\zeta_0 + (2-2/d)\eta_0] / \rho_0$ is the longitudinal kinematic viscosity.
Furthermore, we integrate these spectra over frequency $\omega$ to find the equal-time static correlations:
\begin{align}
    C^{L}_{ab}(\bm{k}) &= \int \frac{d\omega}{2\pi} C^{L}(k, \omega) P^L_{ab} = \frac{k_B T}{\rho_0} \frac{k_a k_b}{k^2}, \label{eq:app_static_L} \\[5pt]
    C^{T}_{ab}(\bm{k}) &= \int \frac{d\omega}{2\pi} C^{T}(k, \omega) P^T_{ab} = \frac{k_B T}{\rho_0} \left(\delta_{ab}- \frac{k_a k_b}{k^2} \right), \label{eq:app_static_T} \\[5pt]
    C_{\rho\rho}(\bm{k}) &= \int \frac{d\omega}{2\pi} C_{\rho\rho}(\bm{k},\omega) = \frac{\rho_0 k_B T}{c_T^2}. \label{eq:app_static_rho}
\end{align}
These results confirm that our model satisfies the equipartition theorem.
The static velocity correlations are determined directly by the equipartition theorem:
\begin{align}
    C_{ab}(\bm{k}) = \frac{k_B T}{\rho_0} \delta_{ab}. \label{eq:app_equip_total}
\end{align}
Our results in Eqs.~(\ref{eq:app_static_L}) and (\ref{eq:app_static_T}) are simply the decomposition of this total correlation.
Notably, due to this equipartition theorem, the velocity correlations $C^L_{ab}(\bm{k})$ and $C^T_{ab}(\bm{k})$ are independent of $c_T$, while the density fluctuations $C_{\rho\rho}$ are suppressed as $c_T^{-2}$.

\subsection{C. Dissipation rate in the large \texorpdfstring{$c_T$}{cT} limit}
The suppression of density fluctuations ($C_{\rho\rho} \propto c_T^{-2}$) suggests that for large $c_T$, we can approximate $\rho \approx \rho_0$.
In this limit, the dynamics of $\psi$ [Eq.~(\ref{eq:app_cd})] simplify to
\begin{align}
    \frac{\partial \psi}{\partial t} + \bm{v} \cdot \nabla \psi = D_0 \nabla^2 \psi - \nabla \cdot \bm{J}_{\mathrm{ran}}.
\end{align}
This equation is formally identical to the advection-diffusion equation in an incompressible fluid.
However, the velocity field $\bm{v}$ still contains longitudinal and transverse modes, and as shown by equipartition, the amplitude of $C^L_{ab}(\bm{k})$ does not vanish.
We must therefore verify that the longitudinal modes do not contribute to the physical quantity of interest, the dissipation rate $\langle \dot{\varepsilon}_{\rm diss}\rangle$.
As shown in the main text, in the steady state, $\langle \dot{\varepsilon}_{\rm diss}\rangle$ is balanced by the work rate, $\langle \dot{\varepsilon}_{\rm work}\rangle$, which is given at the linear level by [Eq.~(\ref{eq:appendix_wr_lin})]
\begin{align}
    \langle \dot{\varepsilon}_{\rm work}\rangle_{\rm lin} = \frac{D_0}{\chi} G^2 + \frac{G^2}{\chi} \int \frac{d^d\bm{k}d\omega}{(2\pi)^{d+1}} \mathcal{G}_0(\bm{k},\omega) C_{yy}(\bm{k},\omega),
    \label{eq:appendix_wr_lin_revisit}
\end{align}
where $\mathcal{G}_0(\bm{k},\omega) = (-i\omega + D_0 k^2)^{-1}$ is the propagator for $\psi$.
The total velocity correlation $C_{yy} = C^{L}_{yy} + C^{T}_{yy}$ splits the calculation into two parts.
The contribution from the transverse modes $C^{T}_{yy}$ yields the standard incompressible result:
\begin{align}
    \langle \dot{\varepsilon}_{\rm work}\rangle_{\rm T} = \frac{k_B T}{\rho_0} \frac{G^2}{\chi(\nu_0 + D_0)} \int \frac{d^d\bm{k}}{(2\pi)^d} \frac{k^2-k_y^2}{k^4}.
\end{align}
The new contribution arises from the longitudinal modes $C^{L}_{yy}$.
We perform the frequency integration in Eq.~(\ref{eq:appendix_wr_lin_revisit}) to obtain an exact analytical result:
\begin{align}
    \langle \dot{\varepsilon}_{\rm work}\rangle_{\rm L} = \frac{k_B T D_0}{\rho_0} \frac{G^2}{\chi} \int \frac{d^d\bm{k}}{(2\pi)^d} \frac{1}{c_T^2 + k^2 D_0(\Gamma_0 + D_0)} \frac{k_y^2}{k^2}.
\end{align}
This expression for the longitudinal contribution $\langle \dot{\varepsilon}_{\rm work}\rangle_{\rm L}$ explicitly contains a $c_T^2$ term.
Consequently, the longitudinal contribution to the dissipation rate vanishes as $c_T \to \infty$:
\begin{align}
    \langle \dot{\varepsilon}_{\rm work}\rangle_{\rm L} \propto \frac{1}{c_T^2} \quad \text{for} \quad c_T \to \infty.
\end{align}
This suppression occurs due to a mismatch in dynamic timescales.
As $c_T$ increases, the characteristic frequency of the longitudinal sound waves ($\omega \sim c_T k$) becomes increasingly high.
This fast timescale effectively decouples the sound waves from the slow diffusive dynamics of the scalar field, which are characterized by $\omega \sim D_0 k^2$.
Therefore, even though the static amplitude of the longitudinal velocity fluctuations remains finite, its contribution to the dynamic observable, such as the dissipation rate, is suppressed and vanishes in the limit.
Thus, we have proved within the linear approximation that in the limit $c_T \to \infty$, the total dissipation rate converges to the purely transverse (incompressible) result.
This result also holds when nonlinear terms are included.
This justifies why the compressible system simulated at large $c_T$ provides an excellent description of the incompressible model.

\end{widetext}
\end{document}